%% file: p2.tex
\pdfoutput=1
\documentclass[11pt,letterpaper]{article}

\usepackage[T1]{fontenc}
\usepackage{newtxtext,newtxmath}
\usepackage[scaled=0.92]{helvet}
\usepackage{microtype}
\usepackage[letterpaper,margin=1in]{geometry}
\usepackage{setspace}
\usepackage[skip=0.7em plus 0.2em minus 0.1em]{parskip}
\usepackage{amsmath}
\usepackage{booktabs}
\usepackage{array}
\usepackage{makecell}
\usepackage{longtable}
\usepackage{graphicx}
\usepackage{float}
\usepackage{xcolor}
\usepackage{enumitem}
\usepackage{cite}
\usepackage{xurl}
\usepackage{titlesec}
\usepackage[font=small,labelfont=bf,labelsep=period,skip=6pt]{caption}
\usepackage{tikz}
\usetikzlibrary{arrows.meta,positioning,fit,calc,backgrounds}
\usepackage[most]{tcolorbox}
\usepackage[hidelinks]{hyperref}
\usepackage[capitalise,noabbrev]{cleveref}

\definecolor{accent}{HTML}{1F3A68}
\hypersetup{colorlinks=true,linkcolor=accent,citecolor=accent,urlcolor=accent,
  pdftitle={No Decision Without a Record: Fail-Closed Evidence Completeness and Its Cost in an Agent Policy Decision Point},
  pdfauthor={Animesh Shaw}}

\titleformat{\section}{\sffamily\bfseries\Large\color{accent}}{\thesection}{0.7em}{}
\titleformat{\subsection}{\sffamily\bfseries\large\color{accent}}{\thesubsection}{0.7em}{}
\titleformat{\subsubsection}{\sffamily\bfseries\normalsize\color{accent}}{\thesubsubsection}{0.6em}{}
\titlespacing*{\section}{0pt}{3.2ex plus 0.6ex minus 0.3ex}{1.4ex plus 0.2ex}
\titlespacing*{\subsection}{0pt}{2.6ex plus 0.5ex minus 0.3ex}{1.0ex plus 0.2ex}
\titlespacing*{\subsubsection}{0pt}{2.2ex plus 0.4ex minus 0.2ex}{0.8ex plus 0.1ex}

\setlist{itemsep=0.35em,topsep=0.3em,parsep=0.15em,leftmargin=1.5em}
\newcommand{\tablestyle}{\centering\small\setstretch{1.0}\renewcommand{\arraystretch}{1.2}\setlength{\tabcolsep}{5pt}}
\newcommand{\figstyle}{\centering\setstretch{1.0}}

\newtcolorbox{glance}{enhanced,colback=accent!4,colframe=accent!55,boxrule=0.5pt,arc=2pt,
  left=8pt,right=8pt,top=5pt,bottom=5pt,before skip=10pt,after skip=12pt,
  title={At a glance},fonttitle=\sffamily\bfseries\small,coltitle=white,colbacktitle=accent!85,
  attach boxed title to top left={yshift=-2mm,xshift=4mm},boxed title style={arc=2pt,boxrule=0pt,
  top=1pt,bottom=1pt},toptitle=3pt}
\newtcolorbox{statement}[1]{enhanced,colback=white,colframe=accent!55,boxrule=0.5pt,arc=2pt,
  left=8pt,right=8pt,top=4pt,bottom=4pt,before skip=9pt,after skip=9pt,
  title={#1},fonttitle=\sffamily\bfseries\small,coltitle=accent,colbacktitle=accent!8,
  attach title to upper={\par\smallskip}}
\newtcolorbox{takeaway}[1][Takeaway]{enhanced,breakable,colback=accent!4,colframe=accent!55,boxrule=0pt,arc=0pt,
  borderline west={2.5pt}{0pt}{accent},left=10pt,right=8pt,top=5pt,bottom=4pt,
  before skip=10pt,after skip=12pt,
  title={#1},fonttitle=\sffamily\bfseries\small,coltitle=accent,colbacktitle=accent!4,
  attach title to upper={\par\smallskip}}

\newcommand{\sys}{Oathpost}

\newcommand{\codeus}{\textunderscore\allowbreak}
\newcommand{\code}[1]{{\let\_\codeus\texttt{#1}}}
\newcommand{\appref}[1]{Appendix~\ref{#1}}

\makeatletter
\def\@medaux#1 #2\relax{#1}
\newcommand{\med}[1]{\expandafter\@medaux#1 \relax}
\def\@rngaux#1 [#2] \relax{#2}
\newcommand{\rng}[1]{\expandafter\@rngaux#1 \relax}
\makeatother
\input{glyphtounicode}
\input{generated/numbers.tex}

\makeatletter
\newcommand{\zenododoi}[1]{%
  \def\zz@arg{#1}\def\zz@ph{TO-BE-ADDED}%
  \ifx\zz@arg\zz@ph
    A copy of the same archive is intended for deposit at Zenodo, which assigns a permanent DOI.%
  \else
    The same archive is also archived at Zenodo, DOI~\href{https://doi.org/#1}{#1}.%
  \fi}
\makeatother

\title{\vspace{-1.2cm}\sffamily\bfseries No Decision Without a Record:\\[0.4ex]
  Fail-Closed Evidence Completeness and Its Cost\\[0.4ex]
  in an Agent Policy Decision Point}
\author{Animesh Shaw\\[0.5ex]
  \small Golden Gate University, San Francisco, CA, USA\\
  \small\texttt{ashaw406@my.ggu.edu}}
\date{}

\begin{document}
\maketitle

\begin{abstract}
\noindent\input{sections/abstract.tex}
\end{abstract}


\input{sections/intro.tex}
\input{sections/background.tex}
\input{sections/related.tex}
\input{sections/property.tex}
\input{sections/method.tex}
\input{sections/results-rq1.tex}
\input{sections/results-rq2.tex}
\input{sections/results-rq3.tex}
\input{sections/discussion.tex}
\input{sections/limits.tex}
\input{sections/data.tex}
\input{sections/conclusion.tex}

\section*{Acknowledgements}
\input{sections/ack.tex}

\bibliographystyle{IEEEtran}
\bibliography{p2}

\appendix
\input{sections/glossary.tex}
\input{sections/appendix-commands.tex}

\end{document}

%% file: generated/numbers.tex
\newcommand{\NFaultRuns}{70}
\newcommand{\NFaultTypes}{13}
\newcommand{\NRequests}{155,946}
\newcommand{\NDecisions}{125,112}
\newcommand{\NRefused}{29,166}
\newcommand{\NNoAnswer}{1,668}

\newcommand{\NMissing}{0}
\newcommand{\NDuplicates}{0}
\newcommand{\NOrphans}{0}
\newcommand{\NRefusedRecorded}{1}
\newcommand{\NUnansRecorded}{287}
\newcommand{\NUnansNotRecorded}{1,381}
\newcommand{\NLate}{4,548}
\newcommand{\NChainsOk}{70/70}
\newcommand{\NOfflineOk}{70/70}

\newcommand{\AsyncChainsOk}{5/5}

\newcommand{\SentNone}{11,139}

\newcommand{\RecoveryPgStop}{1 [1--5]}

\newcommand{\UnansRecPgPause}{31}

\newcommand{\UnansRecNetHold}{45}

\newcommand{\UnansRecNetHoldReset}{2}

\newcommand{\UnansNotRecNetHoldReset}{37}

\newcommand{\UnansRecPgSlowFifty}{149}

\newcommand{\WinEnfDecPgSlowFifty}{173}
\newcommand{\WinEnfTotPgSlowFifty}{1,415}
\newcommand{\WinObsDecPgSlowFifty}{181}
\newcommand{\WinObsTotPgSlowFifty}{1,467}
\newcommand{\UnansRecPgSlowTwoFifty}{56}

\newcommand{\WinEnfDecPgSlowTwoFifty}{135}
\newcommand{\WinEnfTotPgSlowTwoFifty}{1,415}
\newcommand{\WinObsDecPgSlowTwoFifty}{157}
\newcommand{\WinObsTotPgSlowTwoFifty}{1,466}

\newcommand{\UnansRecPdpKill}{4}

\newcommand{\ConnErrPdpKill}{1,232}
\newcommand{\UnansNotRecPdpKill}{1,228}

\newcommand{\RepsPdpKill}{5}

\newcommand{\WinEnfDecRedisKill}{79}
\newcommand{\WinEnfTotRedisKill}{1,382}
\newcommand{\WinObsDecRedisKill}{1,129}
\newcommand{\WinObsTotRedisKill}{1,442}

\newcommand{\MissingPgKillAsync}{7}
\newcommand{\RepsPgKillAsync}{5}

\newcommand{\ControlPfifty}{32 [29--37]}
\newcommand{\ControlPninetynine}{245 [95--390]}

\newcommand{\TputAllowFiveHundred}{58.5 [19.4--63.9]}

\newcommand{\AppendMsLow}{12.9 [12.5--27.1]}
\newcommand{\DecideMsLow}{28.6 [28.2--54.2]}
\newcommand{\AppendShareLow}{45 [44--50]}

\newcommand{\SweepChains}{35/35}

\newcommand{\AppendMsLowAsync}{10.5 [10.5--10.9]}

\newcommand{\AppendMsLowObserve}{13.0 [12.2--13.2]}
\newcommand{\DecideMsLowObserve}{29.2 [27.9--30.6]}

\newcommand{\TputAgentsFiveHundred}{240.9 [239.9--244.2]}

\newcommand{\PfiftyLow}{28 [28--32]}
\newcommand{\PninetyNineLow}{57 [42--538]}

\newcommand{\TputAllowAsyncFiveHundred}{59.6 [41.1--63.9]}
\newcommand{\TputRepsAllow}{5 of 5}
\newcommand{\SatAllow}{87.5 [83.8--91.4]}
\newcommand{\SatRateAllow}{100}
\newcommand{\TputRepsAgents}{3 of 5}
\newcommand{\SatAgents}{273.6 [261.4--274.4]}
\newcommand{\SatRateAgents}{320}

\newcommand{\RunsLossAsync}{5/5}
\newcommand{\RunsLossSync}{0/5}
\newcommand{\FisherP}{0.004}
\newcommand{\FisherOneOver}{1/252}

\newcommand{\AsyncLostBeforeMs}{2}
\newcommand{\AsyncLostAfterMs}{75}
\newcommand{\RuleThree}{2.4\times 10^{-5}}
\newcommand{\AppendSerialPfiftySync}{15.6 [14.5--16.3]}
\newcommand{\RttPfiftySync}{1.11 [0.98--1.32]}
\newcommand{\AppendSerialRateSync}{56 [54--64]}
\newcommand{\AppendSameSixteenRateSync}{77 [74--79]}
\newcommand{\AppendOrgsSixteenRateSync}{215 [198--224]}
\newcommand{\AppendSameFourRateSync}{82 [77--88]}
\newcommand{\LockHeldMsSync}{13.0 [12.7--13.5]}

\newcommand{\AppendSameSixteenRateAsync}{96 [90--99]}

\newcommand{\LockHeldMsAsync}{10.4 [10.1--11.1]}
\newcommand{\FlushMs}{2.5}
\newcommand{\SDKCalls}{17,810}
\newcommand{\SDKRuns}{20}
\newcommand{\SDKViolations}{0}
\newcommand{\SDKDoubleRecorded}{65}
\newcommand{\SDKRedisRuns}{5}
\newcommand{\HealthMaxNonPdp}{235}

\newcommand{\NotEvalRedis}{2,118}

\newcommand{\SlowFiftyDonePerS}{1.09}

\newcommand{\SlowTwoFiftyDonePerS}{0.12}

\newcommand{\LatePct}{3.6}
\newcommand{\UnansRecPct}{17}
\newcommand{\ModelRoundTripsMs}{4.5}
\newcommand{\ModelRestMs}{6.0}
\newcommand{\ScaleRatioSweep}{3.1}
\newcommand{\ScaleRatioAppend}{2.8}
\newcommand{\JsonlBufferedPfifty}{0.001 [0.001--0.007]}
\newcommand{\JsonlFlushPfifty}{0.007 [0.006--0.039]}
\newcommand{\JsonlFsyncPfifty}{0.09 [0.08--1.96]}
\newcommand{\JsonlFsyncRate}{7,696 [514--10,450]}

%% file: sections/abstract.tex
AI agents act through tool calls. A governance layer decides whether each call may run and keeps a tamper-evident record of the decision. Tamper evidence does not imply completeness: a hash chain that has lost its newest entries still verifies. We measure whether such a record stays complete when the system fails. The system studied is a policy decision point (PDP) that commits each decision to a per-tenant hash chain in PostgreSQL before it answers, and answers HTTP 503 instead when it cannot. We state five properties of the record and check them with a client-side oracle that compares every request with the chain. In \NFaultRuns{} runs with \NFaultTypes{} kinds of injected fault (crash, freeze, network partition, slowness, full storage; \NRequests{} requests), the oracle found no returned decision missing from the chain, none recorded twice and no record without a request. A control with asynchronous commit lost \MissingPgKillAsync{} acknowledged decisions in \RunsLossAsync{} runs while every chain still verified. The faults did leave records of decisions that never reached the client in time (\NUnansRecorded{} unanswered, \NLate{} late), so the chain lists decisions, not executions. Through the client library (SDK), under three of the faults, no call of a fail-closed enforcing role ran without a recorded ALLOW. The price is a per-tenant limit: the append holds a row lock and takes about \med{\AppendShareLow}\% of a decision's server time at low load, and one tenant sustained a median of \med{\TputAllowFiveHundred} decisions per second with a 99th-percentile latency of at most 500\,ms. All measurements come from one laptop and describe this design on that machine, not a general capacity.

%% file: sections/intro.tex
\section{Introduction}
\label{sec:intro}

\subsection{The problem}

AI agents increasingly act through \emph{tool calls}: they run a database query, send a web (HTTP) request or write an e-mail on someone's behalf. A \emph{governance layer} sits between the agent and its tools. For each call it decides whether the call may go ahead, and it writes down what it decided and on whose authority. That record is what an operator later shows to an auditor, a customer or a regulator. The EU AI Act, for example, requires high-risk AI systems to allow the automatic recording of events over their lifetime~\cite{euaiact2024}.

Most work on such records asks whether they can be \emph{trusted}: could someone alter or delete them without being noticed? The usual tools are hash chains and signed Merkle trees~\cite{schneier1999secure,crosby2009efficient,rfc6962}. A \emph{hash chain} links each record to the one before it by a cryptographic hash, so that changing an earlier record is detectable. A Merkle tree is a tree of such hashes. This paper asks a different question: is the record \emph{complete}? A hash chain from which the newest entries were lost still verifies, because every remaining link is intact. The verifier cannot tell that something is missing.

Completeness is hardest to keep exactly when the system is under stress. The component that makes the decision, the \emph{policy decision point} (PDP), depends on a database, a cache, a policy engine and a network. Any of them can stop, freeze, slow down or fill up. Two things can then go wrong. If the PDP answers ``allow'' and the record of that answer is lost, the action runs with no evidence. If the PDP records a decision that it never managed to deliver, the record lists an action that may never have run.

Several published designs write audit records asynchronously or let logging fail open~\cite{sasidharan2026cosmos,uchibeke2026oap,opadecisionlogs}. The opposite choice, no answer before a durable record, is a stated requirement in current drafts of the Internet Engineering Task Force (IETF)~\cite{sharif2026aat,gruszka2026receipts} and has been prototyped and crash-tested for agent evidence stores~\cite{beshane2026ackpoint,ghosh2026tracekit,arslan2026janus}. To our knowledge, what has not been reported is a measurement of whether a decision point built this way keeps its records complete when its \emph{own dependencies} fail. Such a measurement needs a checker outside the system, and that checker must be shown to detect loss. It should also show where the cost of the guarantee sits.

\subsection{An example}

Suppose an agent may issue refunds. One call asks to refund 450 dollars. The governance layer allows it, tells the agent to proceed, and the agent issues the refund. A moment later the database that holds the record crashes and loses its last few seconds of writes. When the database restarts, the chain verifies, because it is simply shorter. The refund has no record. An auditor who asks on whose authority the refund was made finds nothing, and nothing in the chain shows that anything is missing.

This example is an illustration, not a measurement. However, the control experiment in \cref{sec:rq1} produces exactly this outcome when the database is configured to acknowledge commits before they are flushed to disk.

\subsection{What this paper does}

We study one PDP, the decision point of \sys{}, a runtime governance layer for agents (\cref{sec:background}). It follows one rule: \emph{no decision without a record}. It writes the event for each decision into a hash chain in PostgreSQL \emph{before} it answers. If it cannot, it answers HTTP 503 (service unavailable) instead. The client then applies the role's \emph{fail mode}: either the call is refused (\code{closed}, also called fail-closed), or it runs and leaves a durable \emph{gap marker}, a record that the call ran without a decision (\code{open\_with\_alert}, also called fail-open). A \emph{tenant} is one customer organisation of the service; each tenant has its own chain.

We define what ``complete'' means as five properties (\cref{tab:props}) and test them under controlled failure. Every request carries a unique operation id, and the client keeps its own record of what it was told. After each run, a checker that we call the \emph{oracle} compares every request with the chain and re-verifies the chain with three independent tools. We inject \NFaultTypes{} kinds of fault into the PDP and its dependencies, repeat each five times under steady load, and include a deliberately weakened control that must make the oracle fire. We repeat the central scenarios end to end through the product's client library, and we measure what the guarantee costs.

\begin{table}[!htbp]
\tablestyle
\caption{The five properties of a complete record that this paper checks. \Cref{sec:property} states them formally.}
\label{tab:props}
\begin{tabular}{@{}>{\raggedright\arraybackslash}p{0.25\linewidth}>{\raggedright\arraybackslash}p{0.69\linewidth}@{}}
\toprule
Property & In plain words\\
\midrule
1. Answer recorded & If a client was given a decision, the chain holds exactly one record of it, with the same outcome.\\
2. Single record & No request has more than one decision record.\\
3. No phantom record & Every decision record belongs to a request that was actually sent.\\
4. Chain intact & The chain verifies: sequence numbers are contiguous and every hash link is correct.\\
5. Fail mode honoured & A role that must fail closed runs a call only after a recorded ALLOW. For an observing role, every call that ran left a decision record or a gap marker.\\
\bottomrule
\end{tabular}
\end{table}

\subsection{Research questions}

\begin{enumerate}[label=\textbf{RQ\arabic*.},leftmargin=3.2em,labelsep=0.5em]
\item \textbf{Completeness under failure.} When the database, the network to it, its storage, the policy engine, the cache or the PDP itself fails, does every decision the PDP returns have exactly one record (the properties answer recorded, single record and no phantom record), and does the chain stay intact?
\item \textbf{Fail modes.} When the record cannot be written, does a call of an enforcing role stay unexecuted, and is a call of an observing role accounted for (the property fail mode honoured)?
\item \textbf{Cost.} What does the guarantee cost? How much of a decision is the durable append, where is the per-tenant ceiling, and how much of it is the durable commit itself?
\end{enumerate}

\subsection{Contributions}

\begin{enumerate}[label=\textbf{C\arabic*.},leftmargin=2.6em,labelsep=0.5em]
\item \textbf{A precise statement of completeness.} Five checkable properties for a decision point, including the three outcomes that the design deliberately allows (\cref{sec:property}).
\item \textbf{An oracle with a positive control.} A client-side checker, and a weakened system on which it must fire. It detects acknowledged-but-lost decisions that chain verification accepts (\cref{sec:method,sec:rq1}).
\item \textbf{A fault-injection measurement.} One implementation, \NFaultRuns{} runs and \NFaultTypes{} fault types, with exact counts of violations (none observed) and of the safe-direction imperfections (\cref{sec:rq1,sec:rq2}).
\item \textbf{A cost measurement.} It places the price of the guarantee in a per-tenant, serialised append, both in isolation and inside the running PDP (\cref{sec:rq3}).
\end{enumerate}

None of the mechanisms is new (\cref{sec:related}); the measurement is. The harness, raw data and recomputation scripts are released with the paper (\cref{sec:data}).

\subsection{Summary of findings}

\begin{enumerate}[label=\textbf{F\arabic*.},leftmargin=2.6em,labelsep=0.5em]
\item \textbf{No decision without a record.} In \NFaultRuns{} runs (\NRequests{} requests), the oracle found \NMissing{} returned decisions missing from the chain, \NDuplicates{} requests with two records and \NOrphans{} records without a request. Every chain verified, also with a standalone verifier that has no database access. For the faults tested, a client that was given an answer could always find its record.
\item \textbf{The oracle sees what verification cannot.} With asynchronous commit, a PostgreSQL crash lost \MissingPgKillAsync{} decisions that clients had already received, with at least one loss in each of the \RepsPgKillAsync{} runs. The durable setting lost none in five runs. Every chain of the asynchronous runs still verified.
\item \textbf{The chain lists decisions, not executions.} \NUnansRecorded{} requests got no answer but were recorded, one request was refused but recorded, and \NLate{} decisions (\LatePct{}\% of all decisions) arrived after the client's 2-second deadline. All of these are in the safe direction for fail-closed roles: a record of a call that did not run, never a call without a record. An auditor therefore cannot read ``recorded'' as ``ran''.
\item \textbf{The fail mode held end to end, for the faults tested.} Through the client library, under a PostgreSQL crash, a PDP kill and a Redis crash (\SDKCalls{} calls), we found \SDKViolations{} violations of the fail-mode property. When the PDP was slow, \SDKDoubleRecorded{} observing calls were recorded twice: once by the client, which had given up, and once by the PDP, which had not.
\item \textbf{The price is a per-tenant ceiling.} At low load, writing the event takes about \med{\AppendShareLow}\% of the server's time per decision. One tenant sustained a median of \med{\TputAllowFiveHundred} decisions per second with a 99th-percentile latency of at most 500\,ms (range \rng{\TputAllowFiveHundred} over five repetitions). Four tenants together completed about \ScaleRatioSweep{} times as many decisions at saturation. A slower database lowers the ceiling sharply.
\end{enumerate}

\subsection{Roadmap}

\Cref{sec:background} defines the terms and describes the system. \Cref{sec:related} reviews related work. \Cref{sec:property} states the property that is tested, and \cref{sec:method} describes the experiments. \Cref{sec:rq1,sec:rq2,sec:rq3} report the results for the three research questions. \Cref{sec:discussion} interprets them, \cref{sec:limits} lists limitations and future work, and \cref{sec:data} says where the data and scripts are. \Cref{sec:conclusion} concludes. \appref{app:glossary} is a glossary of terms and abbreviations.

%% file: sections/background.tex
\section{Background and terminology}
\label{sec:background}

This section defines the terms used in the rest of the paper in plain words, and describes the system that we test. \appref{app:glossary} lists every term and abbreviation alphabetically.

\subsection{Tool calls and governance}

An \emph{AI agent} is a program, usually driven by a language model, that can take actions. It acts by making \emph{tool calls}. A tool call names a tool, for example ``send e-mail'' or ``run SQL query'' (SQL is the standard database query language), and gives it arguments.

A \emph{governance layer} checks each tool call before it runs. The check is made by a \emph{policy decision point} (PDP). A PDP is a service that evaluates the call against a \emph{policy}, a set of rules written by the organisation, and returns a \emph{decision}: \code{ALLOW}, \code{DENY}, or \code{REQUIRE\_APPROVAL} (a human must approve first). The part of the system that asks the PDP and acts on the answer is the \emph{enforcement point}. In this paper the enforcement point is the \emph{SDK} (software development kit), a client library that wraps the tool function so that every call first asks the PDP.

Each decision is written down as an \emph{audit event}: a record that holds the call's input, the result of every rule, the policy version and the call's \emph{operation id}, a unique identifier that lets a request be matched with its record. The \sys{} governance layer studied here can also be reached through an MCP (Model Context Protocol) gateway or an egress proxy, which checks an agent's outbound network requests. Those paths use the same decision function; this paper studies the SDK path.

\subsection{The system studied}

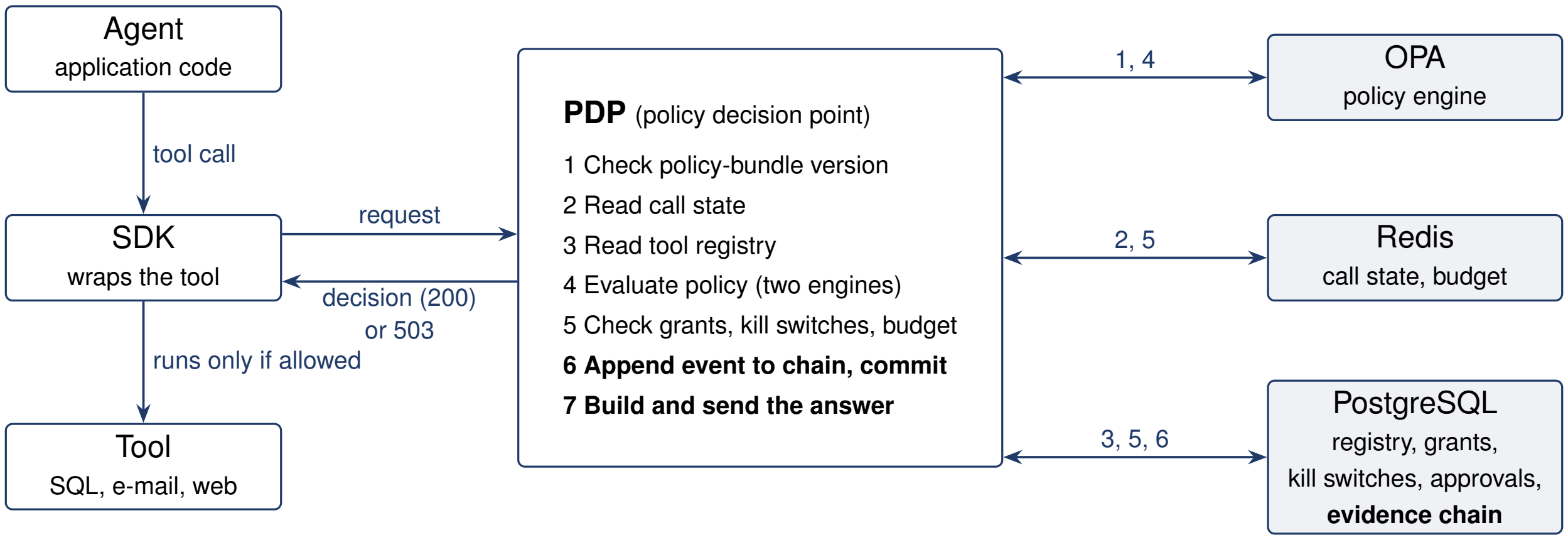
\begin{figure}[!htbp]
\figstyle
\begin{tikzpicture}[
  font=\small\sffamily,
  box/.style={draw=accent,rounded corners=2pt,fill=white,align=center,minimum height=9mm,inner sep=3pt,line width=0.6pt},
  dep/.style={box,fill=accent!7,minimum width=31mm},
  arr/.style={-{Stealth[length=2mm]},line width=0.7pt,accent},
  darr/.style={{Stealth[length=2mm]}-{Stealth[length=2mm]},line width=0.7pt,accent},
  lab/.style={font=\scriptsize\sffamily,align=center,inner sep=1.5pt,fill=white}
]
\node[box,minimum width=29mm] (agent) at (0,2.2) {Agent\\[-1pt]{\scriptsize application code}};
\node[box,minimum width=29mm] (sdk) at (0,0) {SDK\\[-1pt]{\scriptsize wraps the tool}};
\node[box,minimum width=29mm] (tool) at (0,-2.2) {Tool\\[-1pt]{\scriptsize SQL, e-mail, web}};
\node[box,minimum width=51mm,minimum height=44mm,anchor=center,align=left,inner sep=5pt] (pdp) at (6.5,0)
  {\textbf{PDP} {\scriptsize(policy decision point)}\\[3pt]
   \scriptsize 1\ Check policy-bundle version\\
   \scriptsize 2\ Read call state\\
   \scriptsize 3\ Read tool registry\\
   \scriptsize 4\ Evaluate policy (two engines)\\
   \scriptsize 5\ Check grants, kill switches, budget\\
   \scriptsize \textbf{6\ Append event to chain, commit}\\
   \scriptsize \textbf{7\ Build and send the answer}};
\node[dep] (opa) at (13.4,1.9) {OPA\\[-1pt]{\scriptsize policy engine}};
\node[dep] (redis) at (13.4,0) {Redis\\[-1pt]{\scriptsize call state, budget}};
\node[dep] (pg) at (13.4,-2.1) {PostgreSQL\\[-1pt] \scriptsize registry, grants,\\[-1pt] \scriptsize kill switches, approvals,\\[-1pt] \scriptsize\bfseries evidence chain};
\draw[arr] (agent) -- (sdk) node[midway,right=1pt,lab,fill=none]{tool call};
\draw[arr] (sdk) -- (tool) node[midway,right=1pt,lab,fill=none]{runs only if allowed};
\draw[arr] ($(sdk.east)+(0,0.25)$) -- ($(pdp.west|-sdk.east)+(0,0.25)$) node[midway,above=0pt,lab,fill=none]{request};
\draw[arr] ($(pdp.west|-sdk.east)+(0,-0.25)$) -- ($(sdk.east)+(0,-0.25)$) node[midway,below=0pt,lab,fill=none]{\shortstack{decision (200)\\or 503}};
\draw[darr] (opa.west) -- ($(pdp.east|-opa.west)$) node[midway,above=0pt,lab,fill=none]{1, 4};
\draw[darr] (redis.west) -- (pdp.east|-redis.west) node[midway,above=0pt,lab,fill=none]{2, 5};
\draw[darr] (pg.west) -- (pdp.east|-pg.west) node[midway,above=0pt,lab,fill=none]{3, 5, 6};
\end{tikzpicture}
\caption{The parts of the system. The SDK asks the PDP about each tool call. The PDP runs the numbered steps, using OPA, Redis and PostgreSQL, and answers only after step 6 has committed the event.}
\label{fig:arch}
\end{figure}

\Cref{fig:arch} shows the parts of the system. The PDP is a web service. The SDK sends it one HTTP request for each tool call (\code{POST /v1/decide}) and waits for the answer. The PDP uses three other services.

\begin{itemize}
\item \textbf{PostgreSQL} is a relational database. It holds the registry of tools, delegation grants, \emph{kill switches} (flags with which an operator can stop an agent or tool at once), pending approvals and, above all, the evidence chain.
\item \textbf{Redis} is an in-memory store. It holds call state such as rate counters, repeated-call counters and budgets.
\item \textbf{OPA} (Open Policy Agent~\cite{opa}) is a policy engine. The PDP evaluates every policy twice, once in OPA and once in an in-process Python evaluator, and the two must agree on every rule result. If they disagree, the PDP answers 503.
\end{itemize}

\subsection{Roles, modes and fail modes}

Policies are written per \emph{role}, the identity under which an agent acts. Each role has a \emph{mode} and a \emph{fail mode} (\cref{tab:roles}). The mode says what the policy decides. The fail mode says what the \emph{client} does when it gets no answer from the PDP.

\begin{table}[!htbp]
\tablestyle
\caption{Modes and fail modes. The \emph{effective decision} is the decision that takes effect after the mode is applied.}
\label{tab:roles}
\begin{tabular}{@{}>{\raggedright\arraybackslash}p{0.21\linewidth}>{\raggedright\arraybackslash}p{0.73\linewidth}@{}}
\toprule
Setting & What it does\\
\midrule
\code{enforce} & The effective decision follows the policy: \code{ALLOW}, \code{DENY} or \code{REQUIRE\_APPROVAL}. A denied call does not run.\\
\code{observe} & A call that the policy would deny is recorded as \code{decision=DENY} with \code{effective\_decision=ALLOW}, and it runs.\\
\code{closed} & With no answer, the client refuses the call. This is \emph{fail-closed}.\\
\code{open\_with\_alert} & With no answer, the client lets the call run and writes an \code{UNGOVERNED\_ACTION} \emph{gap marker} to a local journal. The journal is flushed to disk before the call proceeds and delivered to the collector later. This is \emph{fail-open}.\\
\bottomrule
\end{tabular}
\end{table}

The product's documentation recommends \code{closed} for every enforcing role and \code{open\_with\_alert} for observing roles. The SDK learns each role's fail mode from the PDP and caches it. A role may make its organisation's default stricter but never looser. The shipped default is \code{open\_with\_alert}, so an enforcing role that is not set to \code{closed} fails open. Everything this paper claims about enforcing roles is for roles configured \code{closed}.

\subsection{The decision path}

For each request the PDP runs these steps in order (the numbers match \cref{fig:arch}). This sequence is also called the decision pipeline.

\begin{enumerate}[label=\arabic*.]
\item Check that this worker's policy-bundle version equals OPA's (one HTTP round trip).
\item Read the call state in Redis, such as rate and repeated-call counters.
\item Resolve the call's target and read the tool registry in PostgreSQL.
\item Evaluate the policy with both engines, OPA and the in-process Python evaluator, which must agree on every rule result.
\item Check delegation grants, kill switches and approvals in PostgreSQL, and the budget in Redis.
\item Append the audit event to the evidence chain and commit.
\item Build the response and send it.
\end{enumerate}

The response is built only after the commit in step 6. Every failure on this path, including a failed append, is answered with HTTP 503 and no decision.

Redis has one exception, which depends on the role. For a role that is not \code{enforce} with \code{closed}, unavailable call state does not refuse the request. The decision is made without it, with the counters treated as zero, and the affected rules are recorded as \code{NOT\_APPLICABLE} with the detail ``not evaluated'' rather than as passed.

\subsection{Limits and time-outs}

The behaviour under overload and failure depends on a few limits, listed here because later results refer to them.

\begin{itemize}
\item \textbf{Admission.} A PDP worker admits at most 64 decision requests at once; it has 32 handler threads. The next request is answered at once with 503 and a \code{Retry-After} header, and nothing is recorded. To its caller this is the same as an unavailable PDP.
\item \textbf{Database.} Every pooled connection has server-side limits: a statement may run 5\,s, a lock wait may last 2\,s and a transaction may sit idle 15\,s. Waiting for a pooled connection is limited to 2\,s.
\item \textbf{OPA and Redis.} Calls to them time out after 1\,s.
\item \textbf{SDK time-out.} Once a decision request is sent, the SDK waits 2\,s (an HTTP client time-out per phase, which in practice bounds the wait for the answer).
\item \textbf{SDK retries.} The SDK retries at most twice, with jittered back-off, and only if the connection was never established or the PDP answered with a 5xx (or 409). It reuses one \code{Idempotency-Key} per call, so a retried request whose first attempt was recorded is answered from the stored copy. A time-out after sending is not retried, because the PDP may already have recorded the decision.
\end{itemize}

\subsection{The evidence chain}

The events of one \emph{tenant} (one customer organisation) form a \emph{hash chain}. Each event stores a hash that covers the event itself and the hash of the event before it. If an earlier event is changed or removed, every later hash stops matching.

Let $e_i$ be the $i$-th event of a tenant and $h_i$ its hash. Then

\begin{equation}
h_i \;=\; \mathrm{SHA\text{-}256}\bigl(\,h_{i-1} \,\Vert\, \mathrm{JCS}(e_i)\,\bigr), \qquad i = 1, 2, \dots, n,
\label{eq:hash}
\end{equation}

where $\mathrm{JCS}(e_i)$ is the event written as canonical JSON (RFC~8785~\cite{rfc8785}, a fixed way of writing JSON so that equal events give equal bytes), $\Vert$ means concatenation, and $h_0$ is a fixed genesis value for the tenant. The events also carry contiguous sequence numbers $1, 2, \dots, n$.

A scheduler periodically seals signed \emph{Merkle checkpoints} over ranges of events, in the style of RFC~6962~\cite{rfc6962}. A Merkle tree is a binary tree of hashes that allows a short proof that an event belongs to a range. An exported chain can be checked by a standalone verifier that needs no database access.

The append of one event is a single PostgreSQL transaction with these steps:

\begin{enumerate}[label=\arabic*.]
\item lock the tenant's \code{chain\_head} row (\code{SELECT ... FOR UPDATE});
\item check that the event id is not already recorded (this makes a retry harmless);
\item seal the event at the next position, that is, compute its hash;
\item insert the event;
\item advance the chain head, and commit.
\end{enumerate}

The row lock serialises the appends of one tenant and is held until the commit. The commit flushes the database's \emph{write-ahead log} (WAL, the log from which PostgreSQL recovers after a crash), so the flush happens inside the locked section. Tenants never wait for each other. This detail explains the cost results in \cref{sec:rq3}.

PostgreSQL's default setting \code{synchronous\_commit=on} makes a commit return only after the WAL record has been flushed to disk. With \code{synchronous\_commit=off} (\emph{asynchronous commit}) the commit returns earlier, and a server crash within the next few hundred milliseconds can lose transactions that were already reported as committed~\cite{pg16asynccommit}.

%% file: sections/related.tex
\section{Related work}
\label{sec:related}

This section groups related work by theme. Each group ends with a short statement of how this paper differs.

\subsection{Tamper-evident and transparency logs}

Hash-chained secure audit logs go back at least to Schneier and Kelsey~\cite{schneier1999secure}. Crosby and Wallach's history tree gives logarithmic membership and incremental proofs~\cite{crosby2009efficient}. Ma and Tsudik's FssAgg authentication makes forward-secure logs compact and names the \emph{truncation attack}, the deletion of the newest entries~\cite{ma2009new}. The Merkle tree and the audit and consistency proofs of Certificate Transparency~\cite{rfc6962,rfc9162} are the template for the checkpoints of the system studied here. In Certificate Transparency a signed certificate timestamp is the log's promise to include an entry within a maximum merge delay, and a broken promise is found afterwards by auditing~\cite[\S3, \S7.3]{rfc6962}.

\begin{takeaway}[How this paper differs]
These schemes make tampering with, deleting or truncating entries \emph{already written} detectable, given a commitment held by someone else (gossip, signed tree heads). They do not make an action depend on its record. The design measured here takes the opposite choice: no allow is returned before its record is durable. An event that was never durably written leaves nothing for a verifier to check, which is the gap that our positive control exposes.
\end{takeaway}

\subsection{Durability and its testing}

Write-ahead logging forces the log to disk at commit before the commit is reported, and group commit shares that flush among several transactions~\cite{mohan1992aries}. PostgreSQL implements group commit through \code{commit\_delay}. Its documentation warns that asynchronous commit ``should not be used if the client will take external actions relying on the assumption that the transaction will be remembered''~\cite{pg16asynccommit}. That is exactly the situation of a decision point. Studies of crash consistency~\cite{pillai2014all,alagappan2016correlated} and of databases under power faults~\cite{zheng2014torturing} show how easily ``acknowledged'' and ``durable'' diverge below the database. The oracle pattern, comparing client-acknowledged operations with the state that survives, is standard in the Jepsen tradition~\cite{kingsbury2020elle} and in crash testers~\cite{mohan2018crashmonkey}. Verifiable ledger databases reach high append rates by batching and group commit~\cite{yang2020ledgerdb}.

\begin{takeaway}[How this paper differs]
We apply the durability-checker pattern to \emph{authorisation evidence} rather than to data, and we report the cost of the durable commit in a running decision point. Batching and group commit are the remedies we discuss (\cref{sec:discussion}), not the design we measure. Our faults stop at the database process; we do not test power loss (\cref{sec:limits}).
\end{takeaway}

\subsection{Recording before acknowledging, under failure}

Four recent preprints are the closest work, and each anticipates part of ours.

\begin{itemize}
\item \textbf{Beshane} argues that a system cannot both guarantee crash-surviving evidence and acknowledge a policy decision before a durability operation. A single-writer prototype issues signed receipts that state whether the record was synchronised. Buffered evidence reaches 27,193 requests/s, and per-record synchronisation about 242 requests/s at a median of 16\,ms, on one host and with component benchmarks rather than fault injection~\cite{beshane2026ackpoint}.
\item \textbf{Tracekit} offers per-tool-class fail modes: a closed class is blocked while its signer is unreachable, and an open class runs while the signer later writes a signed gap record. It acknowledges after write or after fsync, runs an outage matrix (signer down, partition, witness unreachable, storage full) and has a crash test with an oracle very like ours: every acknowledged call must be stored at its acknowledged position, and every unacknowledged call stored or covered by a gap. 30 \code{kill -9} rounds per mode lost nothing, and the authors note that such a test cannot exercise the window that an asynchronous flush leaves open~\cite{ghosh2026tracekit}.
\item \textbf{Janus} makes a step's proposal and verdict durable in a signed hash chain before its effect is released. It is evaluated under crash injection and a long kill-and-recover soak, and reports the added latency of a durable gated step~\cite{arslan2026janus}.
\item \textbf{Xu} injects ten fault classes, including failure to write the evidence record, into an agent tool-action runtime and judges outcomes with an oracle on the target side. The full configuration reaches about one-seventh of the fail-open throughput~\cite{xu2026idempotency}.
\end{itemize}

\begin{takeaway}[How this paper differs]
The faults here are in the decision point's \emph{own dependencies} (its database, the network to it, its storage, its policy engine and its cache) and in the decision point itself. The oracle reconciles every request of an open-loop workload against a multi-tenant chain. A positive control shows that the oracle detects the loss that chain verification accepts. The cost is located inside a running decision point (the per-tenant lock-held window) and not in an isolated writer.
\end{takeaway}

\subsection{Audit records for agents, and standards work}

Several systems record what agents do. None of them measures completeness under failure of the recorder's dependencies.

\begin{itemize}
\item \textbf{NovaFabric} records agent runs as sealed, replayable capsules with a signed export for auditors~\cite{ardebili2026novafabric,ardebili2026novafabric2}. It measures the completeness of its declared event streams in normal operation. It is fail-open by design: capture must never stall or fail the agent, and events are dropped under an explicit loss policy when the collector is unreachable~\cite{ardebili2026novafabric}.
\item \textbf{Soei} keeps a hash-chained, fsync'd ledger of coding-agent sessions as a passive observer. It verifies the ledger at about 126k events/s (42,398 events normalised from 102,233 raw records over 45 sessions) and lists fault injection as future work~\cite{khan2026soei}.
\item \textbf{Agentic Process Evidence} produces signed attestations of agentic processes tied to policy-as-code, without a quantitative evaluation~\cite{kainan2026ape}.
\item \textbf{An authorisation layer for agent tool calls}~\cite{uchibeke2026oap} is fail-closed on the decision but writes its audit record asynchronously after returning it.
\item \textbf{The Agent Flight Recorder} guarantees the integrity of recorded events, but explicitly ``not that all events reach the recorder''~\cite{bindschaedler2026afr}.
\item \textbf{A multi-vendor gateway for large language models (LLMs) built on Envoy} states that ``observability fails open'' and reports that its first asynchronous audit queue lost data silently in production~\cite{sasidharan2026cosmos}. Its overhead figures (a median of 4\,ms and a 99th percentile of 11\,ms added at more than 2,200 requests/s against a synthetic upstream) cover the gateway's request path, not a durable per-decision record, so we do not compare them with our latencies.
\item \textbf{OPA's own decision logs} are buffered in memory, uploaded periodically, and drop the oldest events when the buffer limit is reached~\cite{opadecisionlogs}.
\end{itemize}

The IETF AUDIT effort (a ``birds of a feather'' session, not yet a working group) has several individual drafts that state requirements close to our property.

\begin{itemize}
\item The Agent Audit Trail draft requires that denied and high-risk actions are preceded by a pre-execution record, and its latest revision adds signed completeness statements~\cite{sharif2026aat}.
\item Other drafts record decisions rather than actions, define a ``fail-closed plane'' in which the decision and enforcement points coincide~\cite{bradleyb2026decision}, require that an approval is marked consumed durably before an \code{ALLOW} is released~\cite{gruszka2026receipts}, and record whether the decision point could evaluate its inputs~\cite{gilda2026record}.
\item One draft states the limit that our oracle works around: whether every relevant action produced a record is a question that ``no check of records can answer''~\cite{zhang2026limits}.
\item Another takes the opposite position for observation-only hooks, which it says should generally fail open~\cite{liu2026ep}.
\end{itemize}

On regulation, the EU AI Act requires that high-risk AI systems ``technically allow for the automatic recording of events (logs) over the lifetime of the system'' (Article~12(1)), and that providers and deployers keep those logs for at least six months (Articles~19(1) and~26(6))~\cite{euaiact2024}. A 2026 amendment moved the dates from which these obligations apply to high-risk systems to December 2027 and August 2028~\cite{euaiomnibus2026}. We make no claim that any particular design satisfies the Act.

\begin{takeaway}[How this paper differs]
The drafts specify formats and requirements, and the systems above measure normal operation or declare a loss policy. None of them reports a measurement of record completeness under failure of the recorder's dependencies.
\end{takeaway}

\subsection{Coverage, fault injection and load generation}

guardscope~\cite{toyoda2026guardscope} measures, against kernel-level ground truth, which executed agent actions a runtime control actually checks. It classifies each as checked, bypassed or prevented across eight topology designs, one of which is an uncontrolled baseline. This is the complementary question: an action that bypasses the control produces no evidence however careful the recorder is. Effect binding for agent step-up authorisation~\cite{kodathala2026approve} binds what a human approves to what runs. Chaos engineering~\cite{basiri2016chaos}, network-level fault injection for microservices~\cite{heorhiadi2016gremlin} and service-level fault injection~\cite{meiklejohn2021service} inject failures to test resilience. Studies of network partitions~\cite{alquraan2018analysis} and of fail-slow components~\cite{gunawi2018failslow} motivate our partition and slow-database faults. Our load generation follows the open-system model~\cite{schroeder2006open} and avoids coordinated omission~\cite{wrk2}.

\begin{takeaway}[How this paper differs]
We assume coverage and measure recording. We apply fault injection to a completeness property that is judged by an oracle outside the system under test.
\end{takeaway}

\subsection{What is new here, and what is not}

\begin{itemize}
\item \textbf{Not new:} committing before acknowledging, which is decades old; record-before-answer for agent decisions, which is a stated requirement in the drafts above and is implemented in the prototypes of Beshane, Tracekit and Janus; per-class fail modes with gap records (Tracekit); an oracle that compares acknowledged operations with stored ones under crashes (Tracekit, Xu and the durability-testing literature); and the observation that serialised durable commits run at tens to hundreds per second.
\item \textbf{New:} a measurement of one deployed design that the works we read do not report. \NFaultTypes{} kinds of fault are injected into the decision point's own dependencies and not only into the log writer. Five properties are checked per request against a multi-tenant chain. A positive control shows that the oracle sees loss that chain verification accepts. The frequency of the safe-direction outcomes (recorded but unanswered, late, recorded twice through the SDK) is counted. And the cost is located inside a running decision point.
\end{itemize}

%% file: sections/property.tex
\section{The property studied}
\label{sec:property}

This section states what the PDP promises and what the experiments test. \Cref{sec:promise} gives the promise in plain words, \cref{sec:faultmodel} the assumptions, and \cref{sec:formal} the five properties as formulas. \Cref{sec:allowed} describes three outcomes that the promise deliberately does not exclude.

\subsection{The promise in plain words}
\label{sec:promise}

The PDP promises one thing about its records: \emph{no decision without a record}. If a client was told a decision, a record of that decision exists.

The PDP keeps the promise by \emph{ordering}. It commits the event to the database before it builds the answer (step 6 before step 7 in \cref{fig:arch}). Anything that prevents the commit becomes an HTTP 503 instead of an answer. The promise therefore does not depend on the PDP surviving after it has answered.

Two conditions must hold for the ordering to be enough. First, the commit must be \emph{durable} when it returns: it must survive a crash of the database. Second, the PDP must never write an event for a request it did not receive, and never write two events for one request. The next sections turn these conditions into checkable properties.

\subsection{Fault and threat model}
\label{sec:faultmodel}

We consider \emph{benign} faults: a component stops, freezes, is cut off, slows down or runs out of space. The faults hit the PDP's dependencies (PostgreSQL, the network to it, its storage, OPA and Redis) and the PDP itself. They are injected one at a time on a single node. The PDP, the database and the operator are assumed honest. The client is trusted to record what it was told.

Two things are out of scope.

\begin{itemize}
\item \textbf{A malicious operator.} A malicious operator or database administrator could delete the newest events before a signed checkpoint covers them. In our exports, up to several hundred events at a time were not yet covered by a checkpoint. Detecting such a deletion requires anchoring the chain head outside the operator's control, which this paper does not measure.
\item \textbf{Coverage.} We assume that every governed call goes through the PDP. An action that bypasses the control leaves no evidence however careful the recorder is. Measuring coverage is a separate problem~\cite{toyoda2026guardscope}.
\end{itemize}

\subsection{The five properties}
\label{sec:formal}

To state the properties precisely we need some notation. Let $R$ be the set of requests that clients sent. Each request $r \in R$ carries a unique operation id $\mathrm{op}(r)$, and the PDP stores that id in the event it writes. For each request the client observes exactly one outcome $\mathrm{out}(r)$:

\begin{itemize}
\item $\mathrm{decision}(e)$: an HTTP 200 answer that names the audit event $e$ and an effective decision;
\item $\mathrm{refused}$: an HTTP 503 answer;
\item $\mathrm{unanswered}$: a time-out or a connection error.
\end{itemize}

After a run, let $C$ be the tenant's chain, that is, the sequence of sealed events, and let $D \subseteq C$ be its \emph{decision events}, the events that record a decision. For an event $e$, $\mathrm{op}(e)$ is its operation id and $\mathrm{eff}(e)$ its effective decision. For a request $r$ whose outcome is $\mathrm{decision}(e)$, $\mathrm{told}(r)$ is the effective decision the client was given.

\begin{statement}{Property 1: Answer recorded}
\begin{equation}
\mathrm{out}(r)=\mathrm{decision}(e)\;\Longrightarrow\;
\bigl|\{x\in C : x=e\}\bigr| = 1 \;\wedge\; \mathrm{op}(e)=\mathrm{op}(r) \;\wedge\; \mathrm{eff}(e)=\mathrm{told}(r)
\label{eq:p1}
\end{equation}
If a client was given a decision $e$, then $e$ occurs in the chain exactly once, under the operation id of that request, and it records the same effective decision that the client was told.
\end{statement}

\begin{statement}{Property 2: Single record}
\begin{equation}
\forall r\in R:\quad \bigl|\{e\in D : \mathrm{op}(e)=\mathrm{op}(r)\}\bigr| \le 1
\label{eq:p2}
\end{equation}
For every request, at most one decision event carries its operation id. A request is never recorded twice.
\end{statement}

\begin{statement}{Property 3: No phantom record}
\begin{equation}
\forall e\in D\;\;\exists r\in R:\quad \mathrm{op}(e)=\mathrm{op}(r)
\label{eq:p3}
\end{equation}
Every decision event carries the operation id of some request that was actually sent. The chain contains no decision that nobody asked for.
\end{statement}

\begin{statement}{Property 4: Chain intact}
\begin{equation}
\mathrm{verify}(C)=\mathrm{true}
\label{eq:p4}
\end{equation}
Here $\mathrm{verify}$ is the check that the sequence numbers of $C$ are $1,\dots,n$ without gap or repetition and that \cref{eq:hash} holds for every event. We use the system's own verifier, an SQL check and the standalone verifier (\cref{sec:oracle}).
\end{statement}

\begin{statement}{Property 5: Fail mode honoured}
Let $\mathrm{ran}(r)$ mean that the application executed the tool call of request $r$, and let $\mathrm{gap}(r)$ mean that a gap marker for $r$ exists in the client's journal.
\begin{subequations}\label{eq:p5}
\begin{align}
\text{enforce, closed:}\quad & \mathrm{ran}(r)\;\Longrightarrow\;\mathrm{out}(r)=\mathrm{decision}(e)\;\wedge\;\mathrm{eff}(e)=\mathrm{ALLOW} \label{eq:p5a}\\
\text{observe, open\_with\_alert:}\quad & \mathrm{ran}(r)\;\Longrightarrow\;\bigl(\exists e\in D:\mathrm{op}(e)=\mathrm{op}(r)\bigr)\;\vee\;\mathrm{gap}(r) \label{eq:p5b}
\end{align}
\end{subequations}
For a role that is enforcing and fail-closed, a call runs only if the client received an effective \code{ALLOW}. For an observing role that fails open, every call that runs leaves either a decision event or a gap marker.
\end{statement}

The design aims at the answer-recorded property by ordering: the event is committed before the answer exists. That property then rests on the commit being durable when it returns. With the shipped setting \code{synchronous\_commit=on}, a commit returns after its WAL record is flushed. With \code{off}, it returns earlier, and a crash within the next few hundred milliseconds loses transactions that were already reported as committed~\cite{pg16asynccommit}. The single-record and no-phantom properties depend on the PDP never writing an event for a request it did not receive, and never writing two for one request. Chain integrity depends on the row lock. The fail-mode property is the client's half of the contract.

\subsection{What the property does not promise}
\label{sec:allowed}

Three outcomes are allowed by the design. They are counted in the experiments, not hidden.

\begin{itemize}
\item \textbf{A: refused, but recorded.} The commit succeeded in the database, but the PDP did not learn of it (for example, the acknowledgement was lost). It answered 503 although the record exists.
\item \textbf{B: unanswered, but recorded.} The PDP committed and built its answer, but the client never received it. The PDP may have been killed after the commit, or the client may have given up first.
\item \textbf{C: late.} The answer arrived after the SDK's 2\,s deadline, so an SDK client had already applied the role's fail mode.
\end{itemize}

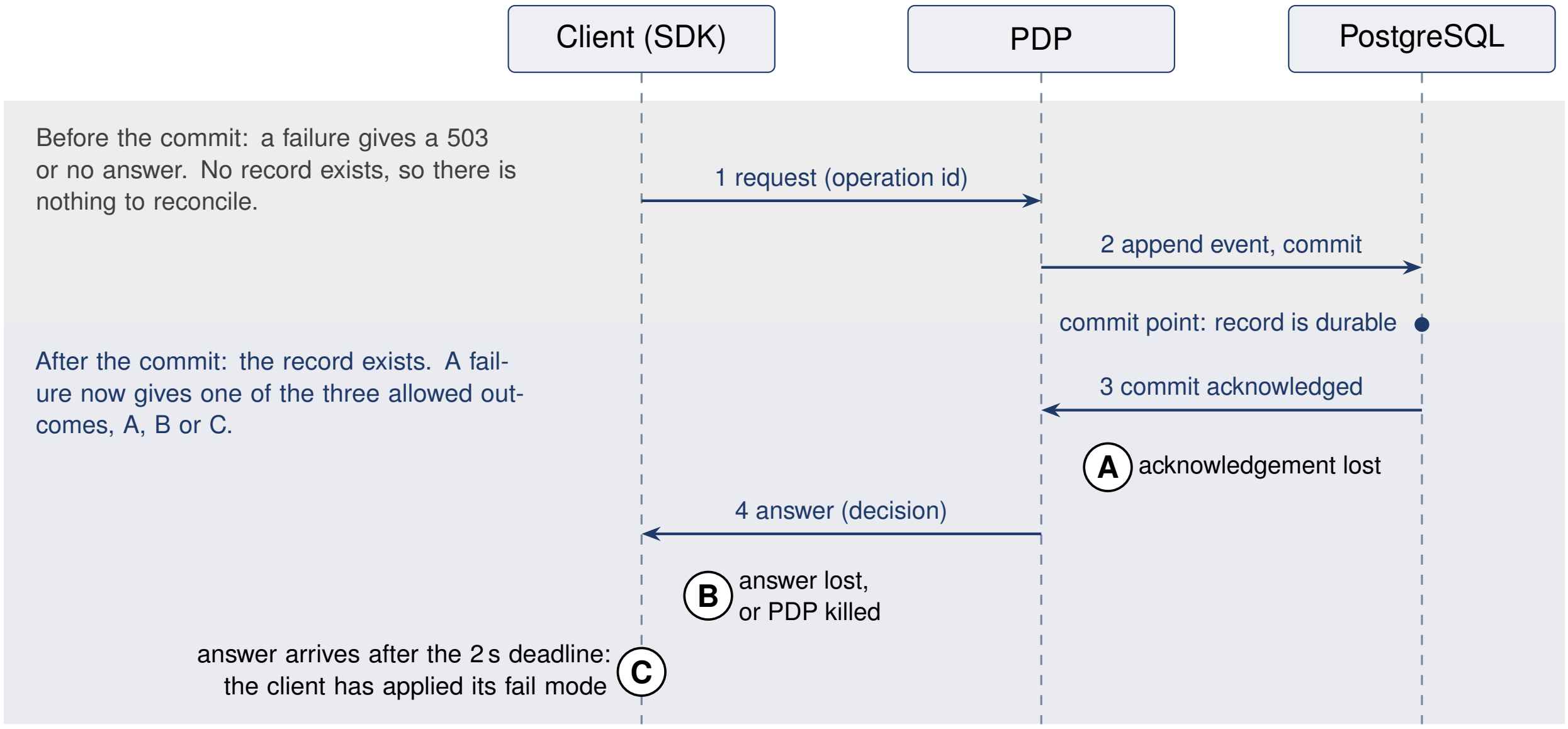
\begin{figure}[!htbp]
\figstyle
\begin{tikzpicture}[
  font=\small\sffamily,
  head/.style={draw=accent,rounded corners=2pt,fill=accent!8,minimum width=28mm,minimum height=7mm,inner sep=2pt},
  life/.style={draw=accent!60,dashed,line width=0.6pt},
  msg/.style={-{Stealth[length=2mm]},line width=0.8pt,accent},
  ml/.style={font=\scriptsize\sffamily,inner sep=1.5pt},
  outc/.style={circle,draw=black,fill=white,inner sep=0pt,minimum size=5mm,font=\small\bfseries\sffamily,line width=0.7pt}
]
\fill[black!7] (-0.1,-0.65) rectangle (16.3,-3.0);
\fill[accent!10] (-0.1,-3.0) rectangle (16.3,-7.2);
\node[font=\scriptsize\sffamily,anchor=north west,text=black!75,text width=5.2cm] at (0.1,-0.8) {Before the commit: a failure gives a 503 or no answer. No record exists, so there is nothing to reconcile.};
\node[font=\scriptsize\sffamily,anchor=north west,text=accent,text width=5.2cm] at (0.1,-3.15) {After the commit: the record exists. A failure now gives one of the three allowed outcomes, A, B or C.};
\node[head] (c) at (6.6,0) {Client (SDK)};
\node[head] (p) at (10.8,0) {PDP};
\node[head] (d) at (14.8,0) {PostgreSQL};
\draw[life] (c.south) -- (6.6,-7.2);
\draw[life] (p.south) -- (10.8,-7.2);
\draw[life] (d.south) -- (14.8,-7.2);
\draw[msg] (6.6,-1.7) -- (10.8,-1.7) node[midway,above=1pt,ml]{1\ request (operation id)};
\draw[msg] (10.8,-2.4) -- (14.8,-2.4) node[midway,above=1pt,ml]{2\ append event, commit};
\fill[accent] (14.8,-3.0) circle (2.2pt);
\node[ml,anchor=east,text=accent] at (14.6,-3.0) {commit point: record is durable};
\draw[msg] (14.8,-3.9) -- (10.8,-3.9) node[midway,above=1pt,ml]{3\ commit acknowledged};
\draw[msg] (10.8,-5.2) -- (6.6,-5.2) node[midway,above=1pt,ml]{4\ answer (decision)};
\node[outc] (A) at (11.5,-4.5) {A};
\node[ml,anchor=west,align=left] at (A.east) {acknowledgement lost};
\node[outc] (B) at (7.3,-5.85) {B};
\node[ml,anchor=west,align=left] at (B.east) {answer lost,\\ or PDP killed};
\node[outc] (C) at (6.6,-6.65) {C};
\node[ml,anchor=east,align=right,text width=5.0cm] at (6.3,-6.65) {\raggedleft answer arrives after the 2\,s deadline: the client has applied its fail mode};
\end{tikzpicture}
\caption{One request in time, running downward. A failure before the commit point leaves no record; after it, A, B and C are the three allowed outcomes of \cref{sec:allowed}.}
\label{fig:windows}
\end{figure}

\Cref{fig:windows} shows where in the flow of one request each outcome arises. All three occur after the commit, when the record already exists.

In all three cases the chain holds a decision for a call that, under \code{closed}, did not run. This is the safe direction: a record of a refused call, never a call without a record. But it means that the chain is not a list of executed actions, and an auditor needs to know that.

The PDP cannot remove outcome B by design. No protocol can make ``committed'' and ``the client knows'' a single atomic step over an unreliable channel. What a system can do is make the client's retry harmless, which is the purpose of the idempotency key (\cref{sec:retries}).

Completeness here is relative to what clients were \emph{told}. The first three properties relate client-visible outcomes to the chain. They do not say that every action that \emph{ran} is recorded; that is the fail-mode property, which needs the application's view and is checked only through the SDK. They do not say that every action an agent attempted reached the PDP either; that is coverage, which we assume.

The oracle has no clock dependence. It matches requests to events by operation id, not by time. Timestamps are used only to classify answers as late and to place requests in a fault window. The generator, the oracle and the PDP share one host clock.

\subsection{Why chain verification alone is not an oracle}
\label{sec:whyverify}

A chain from which the last $k$ events were lost still verifies: every remaining link is intact. This is well known from the tamper-evident logging literature, where truncation is detected only against a commitment held by someone else~\cite{crosby2009efficient,rfc9162}. For loss of the tail, such a commitment would be an external anchor that remembers the chain head. Our oracle instead uses the clients' own record of what they were told. It is therefore built on the client side.

%% file: sections/method.tex
\section{Experimental method}
\label{sec:method}

This section describes the experiments. \Cref{sec:overview} gives the goals and an overview, \cref{sec:env} the test environment, and \cref{sec:load,sec:oracle} the load generator and the oracle. \Cref{sec:faults,sec:control,sec:sdkmethod} describe the fault experiments, and \cref{sec:costmethod} the cost measurements. Every number in this paper is \emph{measured} on the test machine unless it is explicitly labelled \emph{modelled}.

\subsection{Goals and overview}
\label{sec:overview}

The goal is to find out whether the evidence record of a PDP stays complete under failure (RQ1, RQ2), and what the guarantee costs (RQ3). \Cref{tab:experiments} lists the experiments. Each is a controlled run: one factor is varied, everything else is held fixed, and every run is repeated five times.

\begin{table}[!htbp]
\tablestyle
\caption{The experiments, the research question each answers, and what is varied and measured.}
\label{tab:experiments}
\begin{tabular}{@{}>{\raggedright\arraybackslash}p{0.19\linewidth}>{\raggedright\arraybackslash}p{0.07\linewidth}>{\raggedright\arraybackslash}p{0.27\linewidth}>{\raggedright\arraybackslash}p{0.21\linewidth}>{\raggedright\arraybackslash}p{0.17\linewidth}@{}}
\toprule
Experiment & RQ & Varied & Measured & Runs\\
\midrule
Fault matrix & 1 & \NFaultTypes{} fault types, and a no-fault control & Completeness and integrity (properties 1 to 4) & 14 conditions $\times$ 5 runs, 75\,s each\\
Positive control & 1 & PostgreSQL killed, with asynchronous commit & Completeness and integrity (properties 1 to 4) & 5 runs\\
SDK end to end & 2 & PostgreSQL, PDP or Redis killed, or no fault & Fail mode honoured (property 5) & 4 conditions $\times$ 5 runs, 60\,s each\\
Load sweeps & 3 & Offered load; commit mode; one or four tenants & Latency, throughput, append share & 4 cells $\times$ 5 repetitions\\
Append benchmark & 3 & Writers, chains, commit mode & Appends per second, latency & 5 repetitions per mode\\
File-append floor & 3 & Flush mode & Latency & 5 repetitions\\
\bottomrule
\end{tabular}
\end{table}

\subsection{Test environment}
\label{sec:env}

All experiments ran on one laptop. \Cref{tab:env} describes it, with the software versions. The system measured is \sys{} version 0.1.0-dev with its shipped defaults, and one PDP worker.

\begin{table}[!htbp]
\tablestyle
\caption{The test environment. All components run on this one machine and talk over loopback.}
\label{tab:env}
\begin{tabular}{@{}>{\raggedright\arraybackslash}p{0.22\linewidth}>{\raggedright\arraybackslash}p{0.72\linewidth}@{}}
\toprule
Item & Value\\
\midrule
Machine & Laptop, Intel Core i9-14900HX (24 cores, 32 threads), 16\,GB RAM, NVMe SSD\\
Operating system & Windows 11 (build 26300), ``Balanced'' power plan\\
Containers & Docker Desktop 29.7.2, running a WSL2 virtual machine with 32 vCPUs and 4\,GB of memory\\
Database & PostgreSQL 16.15 with TimescaleDB 2.30.1, in a container (\code{fsync=on}, \code{wal\_sync\_method=fdatasync}, \code{full\_page\_writes=on}, \code{max\_connections=300}; \code{synchronous\_commit=on} unless stated)\\
Cache and policy engine & Redis 7.4.9 and OPA 1.22.2, each in a container\\
PDP and tools & On the host: \sys{} 0.1.0-dev (one uvicorn 0.52.1 worker), Python 3.13.13, FastAPI 0.140.0, psycopg 3.3.4, the load generator and the oracle\\
Network & Loopback only; no TLS; PostgreSQL, Redis and OPA reached through forwarded ports\\
Other load & Ordinary desktop use of the machine continued during the runs\\
\bottomrule
\end{tabular}
\end{table}

\subsection{Load generation and per-request records}
\label{sec:load}

The load generator is \emph{open-loop}~\cite{schroeder2006open}. It releases requests on a seeded Poisson timetable (random arrival times at a fixed average rate) whatever the server is doing, and it measures latency from the \emph{intended} send time. A stalled server therefore shows up as latency instead of being hidden (the problem known as coordinated omission~\cite{wrk2}). Each request is released from a timer thread that uses the operating system's high-resolution timer, and the 99th percentile of the scheduling lag was below 2\,ms in every fault run.

For each request the generator stores one record: the operation id (unique per request), the role, the intended and completion times, and the outcome. The outcome is the HTTP status; for a 200 answer also the \code{audit\_event\_id}, \code{decision} and \code{effective\_decision}; for a 503 the reason given in the body; otherwise the client error. A response is awaited for 10\,s. The record also notes whether the answer arrived within the SDK's 2\,s deadline, counted from the actual send time. The client never resends a request.

\subsection{The completeness oracle}
\label{sec:oracle}

After a run, the PDP is hard-killed so that nothing can commit later, and the oracle reads the tenant's chain directly from PostgreSQL. It matches each request with the decision events that carry its operation id, tests the first three properties (answer recorded, single record, no phantom record), and counts the three allowed outcomes of \cref{sec:allowed}.

For chain integrity (property 4) the oracle runs three independent checks. The system's own \code{AuditStore.verify} re-hashes every event. An SQL query checks that the sequence numbers are exactly $1,\dots,n$ with one row per event id. The standalone verifier is run on an export: the chain is written to a JSONL file, and \code{python -m oathpost\_verify verify-chain} and \code{verify-export} are run on that file with no database access. A run counts as clean only if all of these checks pass. \Cref{tab:checks} maps each property to its check.

\begin{table}[!htbp]
\tablestyle
\caption{Each property, the check that tests it, and where the result is reported.}
\label{tab:checks}
\begin{tabular}{@{}>{\raggedright\arraybackslash}p{0.21\linewidth}>{\raggedright\arraybackslash}p{0.55\linewidth}>{\raggedright\arraybackslash}p{0.16\linewidth}@{}}
\toprule
Property & Check & Reported in\\
\midrule
1. Answer recorded & Every returned event id is in the chain, under the request's operation id, with the returned effective decision & \cref{tab:checks-result}\\
2. Single record & No operation id has two decision events; SQL: one row per event id & \cref{tab:checks-result}\\
3. No phantom record & Every decision event's operation id was sent by the client & \cref{tab:checks-result}\\
4. Chain intact & System verifier; SQL check of sequence numbers; standalone verifier on an export & \cref{tab:checks-result}\\
5. Fail mode honoured & SDK runs: execution log of the tool body, compared with the chain and the gap journal & \cref{tab:sdk}\\
\bottomrule
\end{tabular}
\end{table}

\subsection{Faults and workload}
\label{sec:faults}

Each fault run drives a fresh tenant at 30 requests per second for 75\,s. Half of the requests are valid calls by an \code{enforce} role with \code{closed} fail mode. The other half are violating calls by an \code{observe} role with \code{open\_with\_alert}. The fault is injected at 20\,s and healed at 40\,s. The PDP is not restarted unless the PDP itself is the fault. At this rate the no-fault control answers every request (\cref{sec:rq1}), so the faults, not saturation, drive the outcomes. In this workload both roles receive the effective decision \code{ALLOW}: the observing role's violations are recorded as \code{DENY} with effective \code{ALLOW}. An enforcing \code{DENY} is therefore not exercised under fault.

\Cref{tab:faults-def} lists the thirteen faults. 

\begin{table}[!htbp]
\tablestyle
\caption{The thirteen fault types and the no-fault control, with how each is injected.}
\label{tab:faults-def}
\begin{tabular}{@{}>{\raggedright\arraybackslash}p{0.25\linewidth}>{\raggedright\arraybackslash}p{0.49\linewidth}>{\raggedright\arraybackslash}p{0.2\linewidth}@{}}
\toprule
Condition & How it is injected & What it models\\
\midrule
No fault (control) & Nothing & Normal operation\\
\addlinespace[0.3ex]
\multicolumn{3}{@{}l}{\textit{PostgreSQL}}\\
PostgreSQL stopped & \code{docker stop -t 2} (SIGTERM, fast shutdown), then \code{start} & Orderly shutdown\\
PostgreSQL killed & \code{docker kill} (SIGKILL), then \code{start}; crash recovery from the WAL & Database crash\\
PostgreSQL frozen & \code{docker pause} (cgroup freezer): TCP stays up, nothing answers & Hung process\\
\addlinespace[0.3ex]
\multicolumn{3}{@{}l}{\textit{Network between PDP and PostgreSQL (through a TCP proxy)}}\\
Network partition, healed & The proxy holds all bytes in both directions and holds new connections; at the heal it delivers the held bytes in order & Outage shorter than TCP's give-up time\\
Network partition, reset & As above, but at the heal every connection is aborted with RST & Outage longer than TCP's give-up time\\
Slow PostgreSQL & The proxy delays every chunk by 50\,ms or 250\,ms in each direction & Distant or congested database\\
\addlinespace[0.3ex]
\multicolumn{3}{@{}l}{\textit{Evidence storage}}\\
Evidence storage full & The evidence table lives on a 64\,MB RAM-backed (tmpfs) tablespace; at 20\,s a filler file leaves 512\,KB free; at 40\,s it is deleted & Disk full for the evidence table\\
\addlinespace[0.3ex]
\multicolumn{3}{@{}l}{\textit{PDP, policy engine and cache}}\\
PDP killed & \code{taskkill /F} (the Windows counterpart of SIGKILL) at 15, 35 and 55\,s, each followed at once by a restart & Process crash between commit and answer\\
OPA killed; OPA frozen & SIGKILL (OPA restarts with no policy loaded); \code{docker pause} & Policy-engine crash; hang\\
Redis killed; Redis frozen & SIGKILL; \code{docker pause} & Cache crash; hang\\
\bottomrule
\end{tabular}
\end{table}

Two details need explanation. First, the partition and the slowness are produced by a TCP proxy that we wrote for the harness, in the spirit of Toxiproxy~\cite{toxiproxy}. Second, the full-storage fault models ``the evidence table cannot grow''. The WAL and all other tables stay on the normal disk, so it does not model ``the WAL cannot be written''.

Each fault type is repeated five times. The repetitions are interleaved (repetition $k$ of every fault runs before repetition $k+1$ of any), and repetition $k$ uses the same seeded timetable for every fault, so the requests differ only by the fault. The no-fault control is run in the same way.

\subsection{Positive control}
\label{sec:control}

An oracle that never fires might be blind. We therefore repeat the PostgreSQL kill with \code{synchronous\_commit=off}. Under this setting the database acknowledges a commit before its WAL is flushed, and the PostgreSQL documentation states that a crash can then lose transactions that were already reported as committed~\cite{pg16asynccommit}. If the oracle works, it must find decisions that were returned but are not in the chain, while every chain still verifies.

\subsection{End to end through the SDK}
\label{sec:sdkmethod}

The oracle above sees what the PDP answered. It cannot see whether an application ran the action. We therefore also drive the real Python SDK. A tool function is wrapped with the SDK's documented wrapper and called from eight threads (four per role, with a target of 20 calls per second in total) for 60\,s. PostgreSQL, the PDP or Redis is killed at 20\,s and restarted at 40\,s. Each call runs under its own operation id, and the tool body logs that it ran. Afterwards each call is classified: did it run or was it refused, and does it have a decision event, a gap marker in the SDK's journal, or both? This directly tests the fail-mode property (property 5).

\subsection{Cost measurements}
\label{sec:costmethod}

\begin{itemize}
\item \textbf{Load sweeps.} A sweep raises the offered rate step by step (5\,s of warm-up discarded, 20\,s measured) until the service saturates, and then verifies every chain written. There are four cells: valid calls by the enforcing role (``allow''), violating calls by the observing role (``observe''), the allow workload with asynchronous commit, and 100 agents spread over four tenants, each with its own PDP (``agents''). Each cell is repeated five times, interleaved, with the order reversed on alternate repetitions. To locate the cost inside the PDP, we read the PDP's own Prometheus counters before and after every step: the histogram of \code{AuditStore.append} durations and the server-side duration of \code{POST /v1/decide}. The ratio of their sums is the share of server time spent in the append. It includes waiting for the chain-head lock.
\item \textbf{Durability ablation.} The allow sweep and the append benchmark are repeated with \code{synchronous\_commit=off}, switched with \code{ALTER SYSTEM} and a configuration reload on the same container. We do not recommend this setting. It isolates the cost of waiting for the WAL flush inside the chain-head lock.
\item \textbf{Append in isolation.} The product's append benchmark calls \code{AuditStore.append} only, with a decision-sized event and no PDP, OPA or Redis. It is run with one writer on one chain, 4 and 16 writers on one chain, and 4 and 16 writers with one chain each. We also time one client--server round trip (\code{SELECT 1}) on the same path.
\item \textbf{File-append floor.} For scale, the same event is serialised as one JSON line and appended to a local file with no flush, with \code{flush}, and with \code{flush}+\code{fsync} (what the SDK's gap journal does). This floor has none of the properties of the evidence chain and runs on the host file system, not the virtual machine's. It is a scale reference, not a baseline.
\end{itemize}

\subsection{Statistics and reporting}

For each quantity that varies between repetitions we report the median over the five repetitions and the range in square brackets, written ``median [min--max]''. Five repetitions on this machine do not support claims about differences smaller than those ranges, and we make none. Completeness results are exact counts, not estimates. Latencies are percentiles: p50 is the median, and p99 is the value below which 99 of every 100 requests fall.

%% file: sections/results-rq1.tex
\section{Results}
\label{sec:results}

Each of the three subsections below answers one research question. It starts with a summary box, then gives the setup, the result, what the result shows, and a takeaway.

\subsection{RQ1: Is every returned decision recorded?}
\label{sec:rq1}

\begin{glance}
\begin{itemize}
\item \textbf{Question.} When the database, the network, the storage, the policy engine, the cache or the PDP fails, is every decision that the PDP returned in the chain exactly once, and does the chain stay intact?
\item \textbf{How.} \NFaultRuns{} runs of 75\,s at 30 requests per second: \NFaultTypes{} fault types and a no-fault control, five runs each. A client-side oracle checks every request after the run (\cref{sec:method}).
\item \textbf{Observed.} No violations. Every chain verified. A weakened control did lose decisions, and the oracle caught them. The faults also produced records of decisions that never reached the client in time.
\item \textbf{Meaning.} For the faults tested, a client that was given an answer could always find its record. A record does not prove that the call ran.
\end{itemize}
\end{glance}

\subsubsection{Overall outcome}

\Cref{tab:outcomes,tab:checks-result} summarise the \NFaultRuns{} runs. In total the clients sent \NRequests{} requests. Of these, \NDecisions{} received a decision, \NRefused{} a 503 and \NNoAnswer{} no answer (a time-out after 10\,s or a connection error).

\begin{table}[!htbp]
\tablestyle
\caption{What clients received in each condition (measured). Counts are summed over five runs per condition, at 30 requests per second, with the fault active from 20\,s to 40\,s of 75\,s.}
\label{tab:outcomes}
\input{generated/tab-faults-outcomes.tex}
\end{table}

\Cref{tab:outcomes} shows the clients' view. In the control, every one of the \SentNone{} requests received a decision. Under most faults, a fifth to a quarter of all requests, namely those sent while the fault was active, received a 503: the PDP refused because it could not complete its work and record it. Two cases differ. When the PDP itself is killed, requests fail with a connection error while the process is down. When Redis fails, only the enforcing role is refused, as explained in \cref{sec:window}.

\Cref{tab:checks-result} shows what the oracle found. Its columns are:

\begin{itemize}
\item \textbf{Violations:} returned decisions missing from the chain, operation ids with more than one decision event, and decision events without a request (the properties answer recorded, single record and no phantom record).
\item \textbf{Refused, recorded:} answered with a 503, but a decision event exists.
\item \textbf{No answer, recorded:} no answer reached the client, but a decision event exists.
\item \textbf{Answered late:} a decision answered more than 2\,s after the request was sent.
\item \textbf{Chains ok:} the number of runs, out of five, whose chain passed the system's own check and the SQL check (system), or the standalone verifier on an export (offline).
\item \textbf{Recovery:} seconds from the heal action (for the PDP kill, from the restart) to the first one-second interval in which at least 90\% of requests got a decision; median [min--max].
\end{itemize}

\begin{table}[!htbp]
\tablestyle
\caption{What the oracle found in each condition (measured). Counts are summed over five runs per condition.}
\label{tab:checks-result}
\setlength{\tabcolsep}{3.5pt}
\input{generated/tab-faults-checks.tex}
\end{table}

The violations column is zero in every row. The next three columns count the allowed outcomes of \cref{sec:allowed}. They are small, except for answers that arrived late. Every chain passed all three integrity checks: \NChainsOk{} runs by the system's verifier and the SQL check, and \NOfflineOk{} by the standalone verifier on an export. After each fault was healed, service returned without restarting the PDP. The median recovery time was 1\,s in every condition (the resolution is one second). The slowest runs, after PostgreSQL was stopped or killed, took up to 5\,s (range \rng{\RecoveryPgStop}\,s).

In the no-fault control the PDP answered every request at a median latency of \med{\ControlPfifty}\,ms and a p99 of \med{\ControlPninetynine}\,ms (median over the five runs; ranges \rng{\ControlPfifty} and \rng{\ControlPninetynine}\,ms). The faults, not saturation, therefore shaped the outcomes.

\begin{figure}[!htbp]
\figstyle
\includegraphics[width=\linewidth]{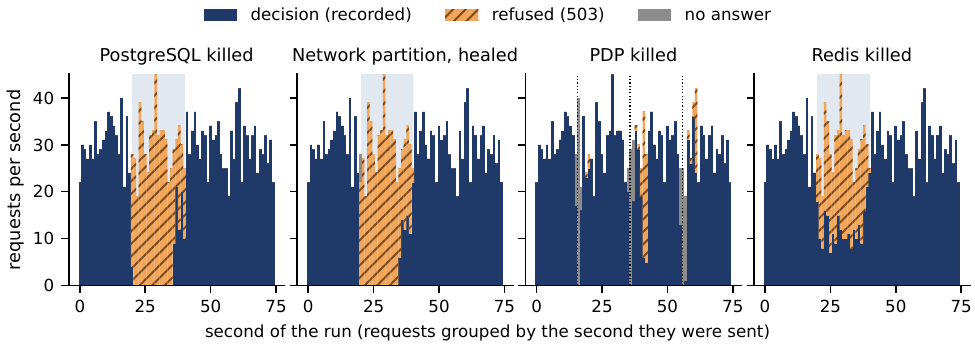}
\caption{Outcome of every request, second by second, in the first run of four faults (measured). The shaded band is the 20-second fault window; dotted lines mark the three PDP kills.}
\label{fig:timeline}
\end{figure}

\Cref{fig:timeline} shows four of the runs. Under the PostgreSQL crash and the partition, the PDP refuses nearly everything for the 20 seconds of the fault and recovers at once. When the PDP is killed, three short gaps appear, one for each kill. Under the Redis crash, the enforcing role is refused while the observing role is still decided; \cref{sec:window} explains why.

\subsubsection{What ``zero violations'' means}

Zero means ``not observed in these runs''. It is not a probability bound. Taken naively as \NDecisions{} independent trials, zero violations would bound the per-decision violation probability by $\RuleThree$ at 95\% confidence (the \emph{rule of three}: three divided by the number of trials). But decisions are not independent trials, and the exposure is concentrated in the few hundred milliseconds around each fault transition. The honest unit is the fault event, of which there are five per fault type (fifteen for the PDP kill). What the data do show is that the oracle can see this kind of violation when it occurs, as the next subsection shows.

\subsubsection{The positive control}

The control repeats the PostgreSQL crash with asynchronous commit. Over \RepsPgKillAsync{} runs, the oracle found \textbf{\MissingPgKillAsync{}} returned decisions that were missing from the chain, with at least one in every run. Each had been answered between \AsyncLostBeforeMs{}\,ms before and \AsyncLostAfterMs{}\,ms after the kill command was issued. The client had been told \code{ALLOW}, or the observing role's recorded \code{DENY}, with an event id, and after crash recovery that event did not exist.

In all of these runs the chain still verified (\AsyncChainsOk{}), because the lost events were the newest and the chain was simply shorter. This is the failure that chain verification cannot see and the client-side oracle can. Loss occurred in \RunsLossAsync{} asynchronous runs and \RunsLossSync{} durable runs. A one-sided Fisher exact test on runs gives $p=\FisherOneOver\approx\FisherP$.

This also shows that the answer-recorded property rests on the durable commit and not only on the ordering. The control exercises the check for that property only. The checks for single record, no phantom record and chain integrity were not given a positive control (\cref{sec:limits}).

\subsubsection{The allowed outcomes, and how often they occur}
\label{sec:allowedcounts}

Over all fault runs, \NRefusedRecorded{} refused request nonetheless had a decision event. \NUnansRecorded{} requests that never got an answer had one (\NUnansNotRecorded{} did not), which is \UnansRecPct{}\% of all unanswered requests. And \NLate{} decisions, \LatePct{}\% of all decisions, were answered after the SDK's 2\,s deadline. The cases concentrate where the design predicts they should:

\begin{itemize}
\item \textbf{PostgreSQL stopped or killed.} Requests in the window were refused within the 2\,s pool wait (the reason is the first database read on the path, the kill-switch store) or shed at the admission bound. When PostgreSQL came back, requests that had been waiting for a pooled connection were decided and recorded seconds after they were sent. These are the \emph{late} answers: more than 2\,s from the moment the request was sent to the moment the answer arrived. The service returned to normal within the recovery times of \cref{tab:checks-result}, without a PDP restart. The PDP's health endpoint answered every probe (four per second) in every run except the PDP kill; the slowest answer took \HealthMaxNonPdp{}\,ms.

The single refused-but-recorded request of the whole matrix is here. An enforcing-role \code{ALLOW} sent 43\,ms after the SIGKILL was issued was committed, and survived crash recovery, but the PDP lost the connection before the commit was acknowledged and answered 503 (``could not be durably recorded''). The chain records an \code{ALLOW} for a call that, under \code{closed}, did not run.
\item \textbf{PostgreSQL frozen, or the network partitioned and healed.} Nothing on the server enforces the statement, lock or idle time-outs while the server is frozen or the bytes are held, so requests already inside a transaction simply waited. The admission bound turned everything else into an immediate 503. On healing, the held transactions committed, some after the client's 10\,s time-out: \UnansRecPgPause{} and \UnansRecNetHold{} no-answer-but-recorded decisions respectively.
\item \textbf{Partition ended by a connection reset.} The reset aborted the held transactions, so the requests without an answer had no event (\UnansNotRecNetHoldReset{} unanswered and not recorded, \UnansRecNetHoldReset{} recorded).
\item \textbf{Slow PostgreSQL.} Adding 50\,ms in each direction (100\,ms per round trip) cut the number of requests sent in the window that received a decision to \WinEnfDecPgSlowFifty{} of \WinEnfTotPgSlowFifty{} for the enforcing role and \WinObsDecPgSlowFifty{} of \WinObsTotPgSlowFifty{} for the observing role. With 250\,ms in each direction the figures were \WinEnfDecPgSlowTwoFifty{} of \WinEnfTotPgSlowTwoFifty{} and \WinObsDecPgSlowTwoFifty{} of \WinObsTotPgSlowTwoFifty{}. The append makes several round trips while it holds the chain-head lock, so requests queued for the lock past its 2\,s time-out were refused as unrecordable (\cref{sec:rq3}). The two slow-database faults produced the most no-answer-but-recorded decisions of all faults (\UnansRecPgSlowFifty{} and \UnansRecPgSlowTwoFifty{} over five runs each).
\item \textbf{Evidence storage full.} Decisions continued until the 512\,KB of headroom was used, which took a few seconds. Then every append failed with ``No space left on device'' and every request was refused as unrecordable. No decision was returned without its event. Deleting the filler file restored service within a second, with no restart of the PDP or the database.
\item \textbf{PDP killed.} \ConnErrPdpKill{} requests failed with a connection error over \RepsPdpKill{} runs (three kills each). \UnansRecPdpKill{} of them had nonetheless been recorded: the PDP died between commit and answer. \UnansNotRecPdpKill{} had not. The restarted PDP needed no repair, because the chain head is in the database and the next append continued the chain.
\item \textbf{OPA killed or frozen.} Requests in the window were refused (OPA unavailable), apart from the few answered just as the fault began or ended. OPA restarts without policy data. Decisions nonetheless resumed within about a second of its restart, without a PDP restart. We attribute this to the PDP's per-request bundle-version check, which re-installs the active bundle (inferred from the code path in \cref{sec:background}, not instrumented).
\item \textbf{Redis killed or frozen.} Here the PDP applies the fail mode itself. Requests by the enforcing, \code{closed} role were refused (\WinEnfDecRedisKill{} of \WinEnfTotRedisKill{} decided in the window of the Redis kill). The observing role was decided without its call-state rules (\WinObsDecRedisKill{} of \WinObsTotRedisKill{}), and those events record the affected rules as \code{NOT\_APPLICABLE} with the detail ``not evaluated'' rather than as passed (\NotEvalRedis{} such events over the Redis runs, all for the observing role). These degraded decisions were slow: each Redis operation waited for its 1\,s time-out, so most of them were answered after the SDK's 2\,s deadline. That is why the Redis faults have the largest counts of late answers. An SDK client would by then have applied \code{open\_with\_alert}, run the call and written a gap marker, while the PDP also recorded a decision for it (measured in \cref{sec:rq2}).
\end{itemize}

\subsubsection{The requests sent during each fault, by role}
\label{sec:window}

\Cref{tab:window} counts what happened to the requests sent while each fault was active, separately for the two roles. A decision for a request sent in the window may have been delivered after the window.

\begin{table}[!htbp]
\tablestyle
\caption{Requests sent while each fault was active, by role (measured; summed over five runs). ``Not answered'' means a 503, a time-out or a connection error.}
\label{tab:window}
\input{generated/tab-window.tex}
\end{table}

For every database, network, storage and policy-engine fault, both roles were refused alike. The exceptions are requests sent just before the heal, which waited and were decided after it, and, for full storage, the requests served from the remaining headroom. The PDP does not distinguish the roles in these cases, because it can record neither. The difference lies in the client, where the role's fail mode decides what the application does (next subsection). Redis is the one dependency whose loss the PDP itself tolerates for roles that are not enforcing and closed.

\begin{takeaway}[Takeaway for RQ1]
In \NFaultRuns{} fault runs the oracle found no returned decision missing, none recorded twice and no record without a request, and every chain verified. The oracle can see loss when it exists: the asynchronous-commit control lost decisions that the chain checks accepted. The faults did produce records of decisions that the client never received in time, so the chain lists decisions, not executions.
\end{takeaway}

%% file: generated/tab-faults-outcomes.tex
\begin{tabular}{@{}lrrrr@{}}
\toprule
Condition (5 runs each) & Requests & Decision & Refused (503) & No answer\\
\midrule
No fault (control) & 11,139 & 11,139 & 0 & 0\\
PostgreSQL stopped & 11,139 & 8,347 & 2,792 & 0\\
PostgreSQL killed & 11,139 & 8,335 & 2,804 & 0\\
PostgreSQL frozen & 11,139 & 8,536 & 2,567 & 36\\
Network partition, healed & 11,139 & 8,517 & 2,573 & 49\\
Network partition, reset & 11,139 & 8,530 & 2,570 & 39\\
Slow PostgreSQL (+50 ms) & 11,139 & 8,609 & 2,321 & 209\\
Slow PostgreSQL (+250 ms) & 11,139 & 8,528 & 2,508 & 103\\
Evidence storage full & 11,139 & 9,303 & 1,836 & 0\\
PDP killed (3 times) & 11,139 & 9,555 & 352 & 1,232\\
OPA killed & 11,139 & 8,304 & 2,835 & 0\\
OPA frozen & 11,139 & 8,412 & 2,727 & 0\\
Redis killed & 11,139 & 9,494 & 1,645 & 0\\
Redis frozen & 11,139 & 9,503 & 1,636 & 0\\
\midrule
All 70 runs & 155,946 & 125,112 & 29,166 & 1,668\\
\bottomrule
\end{tabular}

%% file: generated/tab-faults-checks.tex
\begin{tabular}{@{}lrrrrrrr@{}}
\toprule
Condition & \makecell[r]{\textbf{Violations}} & \makecell[r]{Refused,\\ recorded} & \makecell[r]{No answer,\\ recorded} & \makecell[r]{Answered\\ late} & \makecell[r]{Chains ok,\\ system} & \makecell[r]{Chains ok,\\ offline} & \makecell[r]{Recovery\\ (s)}\\
\midrule
No fault (control) & \textbf{0} & 0 & 0 & 0 & 5/5 & 5/5 & --\\
PostgreSQL stopped & \textbf{0} & 0 & 0 & 213 & 5/5 & 5/5 & 1 [1--5]\\
PostgreSQL killed & \textbf{0} & 1 & 0 & 221 & 5/5 & 5/5 & 1 [1--5]\\
PostgreSQL frozen & \textbf{0} & 0 & 31 & 161 & 5/5 & 5/5 & 1\\
Network partition, healed & \textbf{0} & 0 & 45 & 177 & 5/5 & 5/5 & 1 [0--1]\\
Network partition, reset & \textbf{0} & 0 & 2 & 194 & 5/5 & 5/5 & 1 [0--1]\\
Slow PostgreSQL (+50 ms) & \textbf{0} & 0 & 149 & 316 & 5/5 & 5/5 & 1 [0--1]\\
Slow PostgreSQL (+250 ms) & \textbf{0} & 0 & 56 & 251 & 5/5 & 5/5 & 1 [0--1]\\
Evidence storage full & \textbf{0} & 0 & 0 & 0 & 5/5 & 5/5 & 1 [0--1]\\
PDP killed (3 times) & \textbf{0} & 0 & 4 & 833 & 5/5 & 5/5 & 1 [0--1]\\
OPA killed & \textbf{0} & 0 & 0 & 0 & 5/5 & 5/5 & 1 [0--1]\\
OPA frozen & \textbf{0} & 0 & 0 & 0 & 5/5 & 5/5 & 1 [0--1]\\
Redis killed & \textbf{0} & 0 & 0 & 1,092 & 5/5 & 5/5 & 1 [0--1]\\
Redis frozen & \textbf{0} & 0 & 0 & 1,090 & 5/5 & 5/5 & 1 [0--1]\\
\midrule
All 70 runs & \textbf{0} & 1 & 287 & 4,548 & 70/70 & 70/70 & \\
\bottomrule
\end{tabular}

%% file: generated/tab-window.tex
\begin{tabular}{@{}lrrrr@{}}
\toprule
& \multicolumn{2}{c}{Enforcing role (\texttt{closed})} & \multicolumn{2}{c}{Observing role (open)}\\
\cmidrule(lr){2-3}\cmidrule(lr){4-5}
Condition & Answered & Not answered & Answered & Not answered\\
\midrule
PostgreSQL stopped & 131 & 1,245 & 106 & 1,326\\
PostgreSQL killed & 132 & 1,248 & 105 & 1,335\\
PostgreSQL frozen & 151 & 1,249 & 133 & 1,320\\
Network partition, healed & 140 & 1,274 & 131 & 1,335\\
Network partition, reset & 142 & 1,273 & 139 & 1,327\\
Slow PostgreSQL (+50 ms) & 173 & 1,242 & 181 & 1,286\\
Slow PostgreSQL (+250 ms) & 135 & 1,280 & 157 & 1,309\\
Evidence storage full & 486 & 876 & 471 & 947\\
OPA killed & 21 & 1,359 & 19 & 1,422\\
OPA frozen & 77 & 1,322 & 65 & 1,388\\
Redis killed & 79 & 1,303 & 1,129 & 313\\
Redis frozen & 98 & 1,302 & 1,131 & 321\\
\bottomrule
\end{tabular}

%% file: sections/results-rq2.tex
\subsection{RQ2: Does the fail mode hold end to end?}
\label{sec:rq2}

\begin{glance}
\begin{itemize}
\item \textbf{Question.} When a record cannot be written, does a call of an enforcing, fail-closed role stay unexecuted, and is a call of an observing role accounted for?
\item \textbf{How.} The real Python SDK drives a tool function from eight threads for 60\,s. PostgreSQL, the PDP or Redis is killed at 20\,s and restarted at 40\,s. Four conditions, five runs each: \SDKRuns{} runs and \SDKCalls{} calls.
\item \textbf{Observed.} \SDKViolations{} violations of the fail-mode property. Under the Redis crash, \SDKDoubleRecorded{} observing calls were recorded twice.
\item \textbf{Meaning.} For the faults tested, the fail mode holds from the PDP through the SDK to the application. The double records show a gap between the PDP's and the client's view of time.
\end{itemize}
\end{glance}

\Cref{tab:sdk} shows what the application did with each call, and what evidence exists. In the table, \emph{ran} means that the wrapped tool body executed. \emph{Event} means that a decision event with the call's operation id exists in the chain. \emph{Gap marker} means that an \code{UNGOVERNED\_ACTION} marker is in the SDK's journal. \emph{Both} means event and gap marker; those calls are counted in each of the two preceding columns. Four conditions were run: the control, PostgreSQL killed, the PDP killed once, and Redis killed, at a target of about 20 calls per second from eight threads.

\begin{table}[!htbp]
\tablestyle
\caption{End to end through the Python SDK (measured). Counts are summed over five runs per condition.}
\label{tab:sdk}
\input{generated/tab-sdk.tex}
\end{table}

The invariant checks found \SDKViolations{} violations. No enforcing-role call ran without exactly one recorded \code{ALLOW}. No enforcing-role call ran ungoverned. Every observing-role call that ran had either a decision event or an \code{UNGOVERNED\_ACTION} marker in the SDK's journal, which is flushed to disk before the call proceeds. Enforcing-role calls made during an outage raised \code{GovernanceUnavailable} and did not run.

The table also shows the two costs of the safe direction.

\begin{itemize}
\item \textbf{A refused call can have a decision event.} This happens when the PDP recorded a decision that did not reach the client in time. It occurred once, under the PostgreSQL crash.
\item \textbf{Some calls are recorded twice.} Under the Redis crash, \SDKDoubleRecorded{} observing-role calls over \SDKRedisRuns{} runs have both a decision event and a gap marker. The PDP's degraded decision took longer than the SDK's 2\,s deadline. The SDK therefore applied \code{open\_with\_alert}, ran the call and wrote a gap marker, while the PDP went on to record a decision for the same operation. Neither record is false. But a reader who counts governed and ungoverned actions separately counts these calls in both.
\end{itemize}

\begin{takeaway}[Takeaway for RQ2]
Through the real SDK, under a PostgreSQL crash, a PDP kill and a Redis crash, no enforcing, fail-closed call ran without a recorded \code{ALLOW}, and every observing call that ran left a decision event or a gap marker. The price of the safe direction is a small number of calls that are recorded twice or recorded although refused.
\end{takeaway}

%% file: generated/tab-sdk.tex
\begin{tabular}{@{}llrrrrrrr@{}}
\toprule
& & & \multicolumn{4}{c}{Calls that ran} & \multicolumn{2}{c}{Calls refused}\\
\cmidrule(lr){4-7}\cmidrule(lr){8-9}
Condition & Role & Calls & Total & \makecell[r]{With\\ event} & \makecell[r]{With gap\\ marker} & \makecell[r]{With\\ both} & Total & \makecell[r]{With\\ event}\\
\midrule
No fault (control) & enforce, closed & 3,000 & 3,000 & 3,000 & 0 & 0 & 0 & 0\\
 & observe, open & 3,000 & 3,000 & 3,000 & 0 & 0 & 0 & 0\\
\addlinespace[0.3ex]
PostgreSQL killed & enforce, closed & 2,183 & 1,984 & 1,984 & 0 & 0 & 199 & 1\\
 & observe, open & 1,888 & 1,888 & 1,826 & 62 & 0 & 0 & 0\\
\addlinespace[0.3ex]
PDP killed (once) & enforce, closed & 1,857 & 1,797 & 1,797 & 0 & 0 & 60 & 0\\
 & observe, open & 1,770 & 1,770 & 1,729 & 41 & 0 & 0 & 0\\
\addlinespace[0.3ex]
Redis killed & enforce, closed & 2,123 & 2,018 & 2,018 & 0 & 0 & 105 & 0\\
 & observe, open & 1,989 & 1,989 & 1,989 & 65 & 65 & 0 & 0\\
\bottomrule
\end{tabular}

%% file: sections/results-rq3.tex
\subsection{RQ3: What does the guarantee cost?}
\label{sec:rq3}

\begin{glance}
\begin{itemize}
\item \textbf{Question.} How much of a decision is the durable append, where is the per-tenant ceiling, and how much of the cost is the durable commit itself?
\item \textbf{How.} An append benchmark without the PDP; open-loop load sweeps through the whole PDP for one tenant and for four; the same measurements with asynchronous commit; the slow-database runs; and a file-append floor.
\item \textbf{Observed.} The append takes about \med{\AppendShareLow}\% of the server time per decision at low load. One tenant sustains a median of \med{\TputAllowFiveHundred} decisions per second at a p99 of at most 500\,ms. Four tenants complete about \ScaleRatioSweep{} times as many at saturation. The WAL flush is about \FlushMs{}\,ms of the roughly \med{\LockHeldMsSync}\,ms that each append holds the chain-head lock.
\item \textbf{Meaning.} The price is a limit per tenant, set by what one row lock holds. On this machine that is mostly round trips and client-side work, not the disk flush. A slower database lowers the limit sharply.
\end{itemize}
\end{glance}

\subsubsection{The append in isolation}

The append benchmark calls only \code{AuditStore.append} with a decision-sized event (\cref{sec:costmethod}). \Cref{tab:append} gives, for five combinations of writers and chains, the appends per second and the median latency of one append, for durable and for asynchronous commit.

\begin{table}[!htbp]
\tablestyle
\caption{The append in isolation (measured). Median [min--max] over five repetitions. A \emph{writer} is a thread that appends events; a \emph{chain} is the evidence chain of one tenant.}
\label{tab:append}
\input{generated/tab-append.tex}
\end{table}

One writer on one chain completed \med{\AppendSerialRateSync} appends per second, with a median of \med{\AppendSerialPfiftySync}\,ms per append. One client--server round trip (\code{SELECT 1}) on the same path took a median of \med{\RttPfiftySync}\,ms. More writers on the \emph{same} chain barely help: \med{\AppendSameFourRateSync} appends per second with 4 writers and \med{\AppendSameSixteenRateSync} with 16. Sixteen writers on 16 chains, however, reached \med{\AppendOrgsSixteenRateSync} appends per second. Tenants do not contend, and one chain saturates.

In practical terms, one tenant's chain can accept only about \med{\AppendSameSixteenRateSync} events per second, however many threads write to it. That single-chain ceiling implies that each append holds the chain-head lock for about \med{\LockHeldMsSync}\,ms (one second divided by the ceiling).

What is in those \med{\LockHeldMsSync}\,ms? The locked section contains four round trips (lock, idempotency check, insert, advance the head), the client-side sealing (canonical JSON, SHA-256, model validation; sealing happens after the lock is taken), and the commit with its WAL flush. With asynchronous commit the single-chain ceiling rose to \med{\AppendSameSixteenRateAsync} appends per second, about \med{\LockHeldMsAsync}\,ms held. On this machine, waiting for the WAL flush therefore costs about \FlushMs{}\,ms per append (difference of medians), a minority of the lock-held time. The four round trips take about \ModelRoundTripsMs{}\,ms (four times the measured round trip; \emph{modelled}, not measured directly). That leaves about \ModelRestMs{}\,ms that neither the flush nor the network explains. The code path suggests client-side work, mainly sealing, but we did not time it separately.

\subsubsection{Inside the decision point}

\begin{figure}[!htbp]
\figstyle
\includegraphics[width=\linewidth]{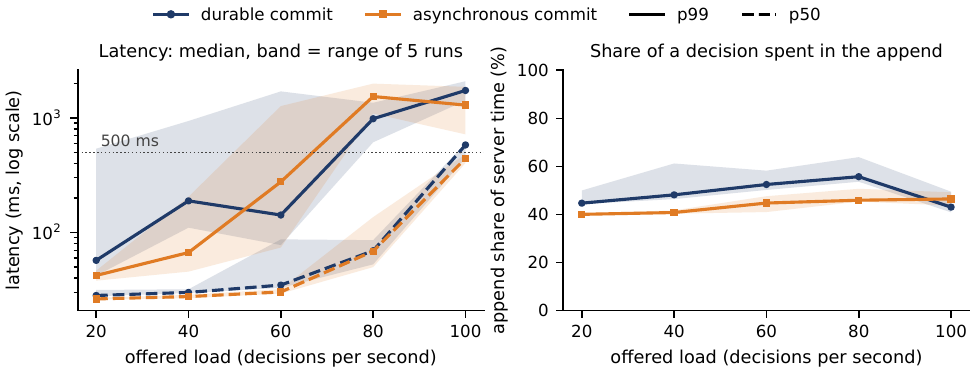}
\caption{One tenant, valid calls by the enforcing role (measured). Left: latency (lines are medians, bands the min--max range). Right: share of server time spent in the append.}
\label{fig:sweeps}
\end{figure}

\Cref{tab:sweeps} and \cref{fig:sweeps} show the open-loop load sweeps. The offered load is raised step by step. \emph{Completed} is the rate of successful decisions. The latencies p50 and p99 are measured from the intended send time. \emph{Append} is the mean time that \code{AuditStore.append} took per decision, and \emph{append share} is its share of the server-side time of \code{POST /v1/decide}, both read from the PDP's own counters over the same step. For the four-tenant cell, the append columns are those of the first tenant's PDP.

\begin{table}[!htbp]
\tablestyle
\footnotesize
\caption{Open-loop load sweeps, one PDP worker per tenant (measured). Median [min--max] over the repetitions that reached each step; their number is in square brackets after the offered load.}
\label{tab:sweeps}
\input{generated/tab-sweeps.tex}
\end{table}

\Cref{fig:sweeps} shows the same sweep for one tenant. In the left panel, the median latency stays close to its low-load value up to 60 offered decisions per second and then rises steeply. The p99 is higher and much noisier, and the bands show how much the repetitions differ. The dotted line marks the 500\,ms bound that we use below. In the right panel, the append share stays around half at every load, for both commit modes. The append share includes the wait for the chain-head lock.

For one tenant on one worker, valid calls by the enforcing role were answered at a median of \med{\PfiftyLow}\,ms at 20 decisions per second, and 99 of every 100 within \med{\PninetyNineLow}\,ms (median p99; the range over the five repetitions is \rng{\PninetyNineLow}\,ms). At that rate the append took a mean of \med{\AppendMsLow}\,ms of a mean server-side decision time of \med{\DecideMsLow}\,ms. The append is therefore \med{\AppendShareLow}\% of a decision. Violating calls by the observing role cost the same at low load: the append took \med{\AppendMsLowObserve}\,ms of \med{\DecideMsLowObserve}\,ms.

The highest throughput with a p99 of at most 500\,ms was \med{\TputAllowFiveHundred} decisions per second (median over \TputRepsAllow{} repetitions; the range is \rng{\TputAllowFiveHundred}, a factor of three, because one repetition already exceeded the bound at 40 offered decisions per second). From about 80 offered decisions per second the service began to shed load: some requests were answered with 503. The decision point completed up to \med{\SatAllow} decisions per second at the \SatRateAllow{} per second step. That is somewhat more than the single-chain ceiling of the benchmark. Towards saturation the append share stays at about half. Waiting for the chain-head lock is counted in the append, and the rest of the time is queueing for threads and the other steps of the pipeline.

In user terms, a tenant whose agents together ask the PDP for more decisions per second than this would see decision latency grow quickly on this machine, unless the load is spread over more tenants.

The tails were noisy. The p99 at a fixed rate varied by up to an order of magnitude between repetitions (\cref{tab:sweeps}). Because of this noise we report throughput at a p99 bound and make no claim from differences inside the ranges.

\subsubsection{Durable versus asynchronous commit}

With asynchronous commit the low-load append fell only from \med{\AppendMsLow}\,ms to \med{\AppendMsLowAsync}\,ms. Throughput at a p99 of at most 500\,ms was \med{\TputAllowAsyncFiveHundred} decisions per second, inside the range of the durable cell. On this test machine, the flush is not what sets the per-tenant ceiling. The round trips and the client-side work held under the lock are. Every chain written by the sweeps verified (\SweepChains{} chains), including steps driven past saturation.

\subsubsection{Many tenants}

Four tenants on four PDPs (100 agents) completed \med{\SatAgents} decisions per second at the \SatRateAgents{} per second step. One tenant completed \med{\SatAllow} at the \SatRateAllow{} per second step. The ratio is about \ScaleRatioSweep{}, consistent with the append benchmark, where 16 writers on 16 chains completed \ScaleRatioAppend{} times as many appends per second as 16 writers on one chain. The ceiling is per tenant, not per deployment.

The p99 bound gives a weaker reading here. Only \TputRepsAgents{} repetitions had any step with a p99 of at most 500\,ms (\med{\TputAgentsFiveHundred} decisions per second in those). The four-tenant p99 was not monotonic in the offered load: the 80 and 160 per second steps had a worse p99 than the 240 per second step in every one of the five repetitions, and the reason is unknown. The four PDPs also shared one host's CPU and memory with their load generator, so this cell measures less than four independent deployments would.

\subsubsection{A slower database lowers the ceiling}

The slow-database faults show the other side of the same lock. From the code, the lock is held from the \code{SELECT ... FOR UPDATE} to the commit, across four client--server round trips before the commit itself. With 100\,ms added per round trip, that is at least 0.4\,s per append, so at most about 2.5 appends per second per tenant (\emph{modelled} from the code, not measured). With 500\,ms added per round trip, the bound is about 0.5 per second.

The measured rates are below these bounds. While the fault was active, the PDP delivered \SlowFiftyDonePerS{} decisions per second to clients with 100\,ms added per round trip (50\,ms in each direction), and \SlowTwoFiftyDonePerS{} per second with 500\,ms added, while 30 requests per second were offered. These rates count decisions whose answer reached the client during the 20-second window. They fall steeply as the added delay grows, in the direction that the lock model predicts.

A deployment that puts the decision point and its database in different availability zones, a few milliseconds apart, should expect its per-tenant ceiling to be set by that latency rather than by the disk. As an illustration (\emph{modelled}): adding 3\,ms to each of the four round trips would add about 12\,ms to the roughly \med{\LockHeldMsSync}\,ms that the lock is held, which roughly halves the ceiling.

\subsubsection{A file-append floor}

For scale, appending the same event as one JSON line to a local file cost a median of \med{\JsonlBufferedPfifty}\,ms without a flush, \med{\JsonlFlushPfifty}\,ms with \code{flush}, and \med{\JsonlFsyncPfifty}\,ms with \code{flush}+\code{fsync} (\med{\JsonlFsyncRate} appends per second; host NTFS, not the virtual machine's disk; measured). The durable, chained, ordered append costs far more than the bare \code{fsync}'d line. But the line has no ordering across writers, no chain, no idempotency and no query path, and nothing in it makes the answer wait for it. It is a scale reference, not a baseline.

\begin{takeaway}[Takeaway for RQ3]
The price of the guarantee is a per-tenant ceiling. A decision spends about \med{\AppendShareLow}\% of its server time in an append that holds one row lock. At about \med{\LockHeldMsSync}\,ms per append, one tenant's chain accepts roughly \med{\AppendSameSixteenRateSync} events per second, and tenants scale independently. On this machine the lock is held mostly for round trips and client-side work; the disk flush is a minority. A slower database lowers the ceiling sharply.
\end{takeaway}

%% file: generated/tab-append.tex
\begin{tabular}{@{}lllll@{}}
\toprule
& \multicolumn{2}{c}{Durable commit} & \multicolumn{2}{c}{Asynchronous commit}\\
\cmidrule(lr){2-3}\cmidrule(lr){4-5}
Writers and chains & Appends/s & p50 (ms) & Appends/s & p50 (ms)\\
\midrule
1 writer, 1 chain & 56 [54--64] & 15.6 [14.5--16.3] & 70 [61--72] & 12.9 [12.5--14.9]\\
4 writers, 1 chain & 82 [77--88] & 45.7 [44.2--48.4] & 99 [96--106] & 37.4 [36.4--38.5]\\
16 writers, 1 chain & 77 [74--79] & 196.3 [186.9--213.0] & 96 [90--99] & 161.8 [144.9--167.3]\\
4 writers, 4 chains & 196 [189--208] & 19.9 [16.9--20.8] & 197 [191--207] & 20.0 [19.0--20.3]\\
16 writers, 16 chains & 215 [198--224] & 71.1 [64.8--78.1] & 215 [191--222] & 70.5 [69.5--81.0]\\
\bottomrule
\end{tabular}

%% file: generated/tab-sweeps.tex
\begin{tabular}{@{}rlllll@{}}
\toprule
\makecell[r]{Offered\\ (/s) [reps]} & \makecell[l]{Completed\\ (/s)} & \makecell[l]{p50\\ (ms)} & \makecell[l]{p99\\ (ms)} & \makecell[l]{Append\\ (ms)} & \makecell[l]{Append share\\ (\%)}\\
\midrule
\multicolumn{6}{@{}l}{\textit{Enforcing role, valid calls, durable commit (one tenant)}}\\[0.2ex]
20 [5] & 19.6 [19.4--20.6] & 28 [28--32] & 57 [42--538] & 12.9 [12.5--27.1] & 45 [44--50]\\
40 [5] & 40.2 [38.5--41.1] & 30 [30--32] & 189 [110--939] & 16.6 [14.6--43.3] & 48 [46--61]\\
60 [5] & 59.6 [58.5--63.9] & 35 [33--87] & 142 [78--1,702] & 19.8 [17.9--135.1] & 52 [50--58]\\
80 [4] & 80.5 [79.5--81.0] & 69 [53--86] & 983 [612--1,355] & 89.3 [55.1--101.4] & 56 [54--64]\\
100 [4] & 87.5 [83.8--91.4] & 580 [447--600] & 1,734 [1,415--2,087] & 232.4 [222.8--244.9] & 43 [41--49]\\
\midrule
\multicolumn{6}{@{}l}{\textit{Enforcing role, valid calls, asynchronous commit (one tenant)}}\\[0.2ex]
20 [5] & 19.6 [19.4--20.6] & 26 [26--27] & 42 [38--46] & 10.5 [10.5--10.9] & 40 [40--40]\\
40 [5] & 40.2 [38.5--41.1] & 28 [27--29] & 67 [46--205] & 11.6 [10.9--13.2] & 41 [40--42]\\
60 [5] & 60.7 [58.5--63.9] & 30 [29--32] & 277 [74--1,268] & 15.8 [13.7--40.6] & 45 [41--48]\\
80 [5] & 77.5 [75.3--79.2] & 69 [50--136] & 1,536 [1,091--1,990] & 97.5 [83.4--114.5] & 46 [45--51]\\
100 [2] & 91.2 [83.3--99.2] & 442 [404--481] & 1,292 [718--1,866] & 188.4 [185.4--191.5] & 46 [44--49]\\
130 [1] & 102.8 & 555 & 1,724 & 210.3 & 37\\
\midrule
\multicolumn{6}{@{}l}{\textit{Observing role, violating calls, durable commit (one tenant)}}\\[0.2ex]
10 [5] & 9.8 [9.6--10.4] & 28 [27--30] & 48 [40--57] & 13.0 [12.2--13.2] & 44 [43--44]\\
20 [5] & 19.6 [19.4--20.6] & 28 [28--29] & 48 [41--56] & 12.2 [12.1--13.0] & 44 [43--44]\\
40 [5] & 40.2 [38.5--41.1] & 31 [30--34] & 268 [57--623] & 15.7 [14.2--42.8] & 47 [46--62]\\
60 [5] & 60.7 [57.4--63.9] & 36 [33--47] & 531 [88--1,332] & 29.9 [18.6--109.9] & 53 [51--58]\\
\midrule
\multicolumn{6}{@{}l}{\textit{Four tenants, 100 agents, durable commit}}\\[0.2ex]
80 [5] & 81.0 [80.0--85.0] & 29 [28--32] & 1,091 [815--2,012] & 74.9 [52.0--131.2] & 62 [55--68]\\
160 [5] & 153.2 [152.4--161.6] & 51 [36--324] & 2,317 [2,022--2,540] & 170.8 [119.6--209.9] & 46 [45--50]\\
240 [5] & 240.9 [235.4--246.2] & 47 [43--56] & 339 [212--811] & 80.8 [32.0--94.9] & 61 [45--68]\\
320 [5] & 273.6 [261.4--274.4] & 924 [884--966] & 1,554 [1,364--1,766] & 329.6 [321.9--354.1] & 38 [35--39]\\
\bottomrule
\end{tabular}

%% file: sections/discussion.tex
\section{Discussion}
\label{sec:discussion}

This section interprets the results. It explains what the ordering of commit before answer buys, where it leaves gaps, what the cost means, and what a practitioner can take from the measurements.

\subsection{What commit-before-answer buys, and what it does not}

Ordering the durable commit before the answer turns the hard direction of completeness, no action without a record, into a local property of one code path. The response object does not exist until the commit has returned, and every exception on the way becomes a 503. The faults could therefore produce only the other direction: records without a client-visible decision. They did.

That direction cannot be closed by the PDP alone. Between ``the database has committed'' and ``the client has the answer'' there is always a window in which a crash, a partition or a client time-out leaves the two disagreeing. An auditor reading the chain must therefore read a decision event as ``this was decided and recorded'', not as ``this ran''. For an enforcing role, whether the call ran is known only to the client.

\subsection{The deadline gap}

The most frequent imperfection was not a lost answer but a \emph{late} one. After an outage, requests that had waited in the connection pool's reconnect queue were decided and recorded seconds later, past the SDK's 2\,s deadline. An SDK client had already applied the fail mode.

\begin{itemize}
\item Under \code{closed}, the call did not run, although the chain records an \code{ALLOW}. The SDK runs contain one such call (\cref{sec:rq2}); the stored data do not say whether it was late or refused after the commit.
\item Under \code{open\_with\_alert}, the call runs and writes a gap marker. The same call is then recorded both as governed and as ungoverned. We measured this \SDKDoubleRecorded{} times under the Redis crash.
\end{itemize}

The PDP does not know the client's deadline. Two remedies follow. Neither was part of the measured version, and neither is evaluated here.

\begin{itemize}
\item \textbf{Propagate the deadline.} The client sends its deadline with the request. The PDP refuses (503, nothing recorded) once the deadline has passed before the append commits. This narrows the window to the last steps of the append.
\item \textbf{Link the gap marker.} The client links its gap marker to the decision it gave up on. An auditor can then tell a recorded-but-abandoned decision from an executed one. This makes the remaining window visible.
\end{itemize}

\subsection{Retries and duplicates}
\label{sec:retries}

A 503 that is sent after the commit has in fact succeeded invites a retry, and the retry could record the same call twice. The SDK's \code{Idempotency-Key} handles the common case, where the first attempt completed and its answer was stored.

Does the key also protect against a 503 whose cause was a lost commit acknowledgement? We checked this by reading the measured code, and it was not observed in any run. By the code, it does not protect. The key's claim is completed in a separate transaction after the append, and any exception releases the claim. Because the event's id is minted per evaluation rather than per call, the retry is a fresh evaluation with a fresh id. The same holds when a secondary event appended after the primary one fails, since the primary has already committed.

The trigger did occur once in our runs (\cref{sec:allowedcounts}): a commit that was made durable just before a database crash was answered with 503. An SDK client that retried that 503 after the database returned would, by this reading, have produced a second record of one call. Completing the idempotency record inside the append transaction closes this gap, at the cost of one more statement under the lock.

\subsection{Blocking versus refusing}

The server-side time-outs (statement, lock, idle in transaction) are the design's mechanism for turning a slow dependency into a prompt 503. They do not fire when the database process is frozen or the network silently holds bytes, because nothing runs on the server to enforce them. In those faults the admission bound did the work. In-flight requests piled up to the limit and the rest were refused at once, so the PDP stayed responsive to health checks and refused rather than queued. A client-side socket time-out on the database connection would bound the frozen case too.

\subsection{Where the cost comes from, and how to lower it}

The append holds the tenant's chain-head lock from the \code{SELECT ... FOR UPDATE} to the commit. Everything inside that window is serialised per tenant: four client--server round trips, the client-side sealing and the WAL flush. On our test machine the flush was about a fifth of the lock-held time (\cref{sec:rq3}), and with a slower network to the database the round trips dominate even more. Tenants do not contend, so aggregate throughput scales with the number of tenants.

Two changes would lift the per-tenant ceiling without weakening the answer-recorded property. Neither is measured here.

\begin{itemize}
\item \textbf{One round trip.} Do the locked part in a single statement or server-side function that locks, checks, inserts and advances the head.
\item \textbf{Group commit.} Seal the event under the lock, release the lock, and answer only after a shared flush that covers the event. PostgreSQL's own group commit cannot help while the lock is held through the commit.
\end{itemize}

Both keep commit-before-answer. Splitting a tenant's chain into several chains, joined by a periodic Merkle root, would also work, at the cost of a more involved verifier.

\subsection{Practical guidance}

The following points follow from the measurements. They are limited to the faults and the version tested.

\begin{itemize}
\item \textbf{Configure enforcing roles as fail-closed.} The shipped default fail mode is \code{open\_with\_alert}. The guarantee for enforcing roles holds only for roles set to \code{closed}.
\item \textbf{Keep synchronous commit on.} The asynchronous-commit control lost decisions that clients had been given, and every chain still verified.
\item \textbf{Read the chain as decisions, not executions.} If an audit needs to know that a call ran, combine the chain with the application's own log or with the gap markers.
\item \textbf{Plan capacity per tenant.} The append is serialised per tenant. Keep the PDP and its database close together, because added round-trip time lowered the rate sharply in our slow-database runs, and spread load over tenants where possible.
\item \textbf{Check completeness from the client side.} A completeness check needs a record held outside the system, and it should be shown to fire on a known failure, as the positive control does here.
\end{itemize}

%% file: sections/limits.tex
\section{Limitations and future work}
\label{sec:limits}

\subsection{Limitations}

Six limitations matter for how far the results can be trusted. Each is followed by one way to address it.

\begin{itemize}
\item \textbf{One machine.} Every component ran on one Windows laptop that was also used for other work. The database ran in a virtual machine, and there was no real network, no TLS and no separate database host. The completeness counts do not depend on speed, but the rates of the safe-direction outcomes and all latency and throughput figures do. \emph{Remedy:} repeat the fault matrix on Linux with the database on a separate host.
\item \textbf{Benign faults and process-level durability.} Faults were benign, injected one at a time on a single node, and they stop the database \emph{process}. We did not test power loss, a failing virtual disk or a malicious operator, who could remove events that no externally anchored checkpoint covers. \emph{Remedy:} anchor chain heads outside the operator's control, and add storage-level fault injection.
\item \textbf{Scope of the path and the workload.} We drove only the SDK's decision path, with one PDP worker, one tenant per fault run, 30 requests per second and one 20-second fault at a time. The workload produced no enforcing \code{DENY}. The fail-mode property was checked under three faults only, and the gap journal was not tested under a crash of the client. \emph{Remedy:} extend the matrix to the MCP gateway and egress paths, multiple workers, combined faults, \code{DENY} and approval outcomes, and a client crash.
\item \textbf{Trust in the oracle.} The positive control covers the answer-recorded property only. The author built both the system and the oracle, and the oracle trusts the client's record and the stored operation id. The chain rows were not stored, so the archive allows re-checking counts and verdicts but not re-hashing events. \emph{Remedy:} add mutation controls for the single-record, no-phantom and chain-intact properties, store chain exports with the data, and have an independent party replicate the study.
\item \textbf{Statistical power.} Each condition was repeated five times, so ``zero violations'' means ``not observed'', and tail latencies vary by up to an order of magnitude between repetitions. The four-tenant p99 was not monotonic and is undiagnosed. \emph{Remedy:} more repetitions on a quiet machine, with confidence intervals, and a diagnosis of the four-tenant tail.
\item \textbf{No baseline, and a superseded version.} We did not measure the common design that answers first and records asynchronously, so cost is reported in absolute terms and against an asynchronous-commit variant. The measured version lacks the two remedies of \cref{sec:discussion}, deadline propagation and in-transaction idempotency. \emph{Remedy:} measure the asynchronous-audit design under the same faults, and re-run the released matrix against newer versions.
\end{itemize}

\subsection{Future work}

Beyond these points, the natural next steps are the two changes of \cref{sec:discussion} that would lift the per-tenant ceiling, a one-round-trip append and group commit, and measuring deadline propagation and gap-marker linking with the same oracle and fault matrix.

%% file: sections/data.tex
\section{Data and artifact availability}
\label{sec:data}

The data and scripts needed to check every number in this paper are distributed with it. The measured version of the system is \sys{} 0.1.0-dev (\cref{tab:env}); it is recorded in every result file.

\subsection{What is provided}

One archive, \code{p2-artifact.tar.gz}. It contains:

\begin{itemize}
\item the measurement harness: the load generator, the completeness oracle, the TCP fault proxy, the fault driver, the SDK check, and the sweep and benchmark drivers;
\item the raw results: one JSON document per run, with the environment, the fault timeline, host samples, oracle counts and verifier verdicts;
\item every client record of every fault run;
\item the scripts that regenerate every table, figure and number of this paper from the stored data, and an independent re-check script.
\end{itemize}

\subsection{Where it is}

The archive is an ancillary file of the arXiv submission: it is the \code{anc/} folder of the upload, listed on the abstract page under ``Ancillary files''. \zenododoi{TO-BE-ADDED}

\subsection{How to recompute the tables}

After unpacking the archive, run two commands from its top directory:

\begin{small}
\begin{verbatim}
python harness/analyze.py            # rewrites results/tables, paper/generated
python harness/independent_check.py  # re-checks all 75 fault runs
\end{verbatim}
\end{small}

The first command regenerates the numbers, tables and figure inputs of this paper byte for byte. The second re-derives the oracle's counts from the stored per-request records without sharing code with the oracle, and should end with the line \code{75/75 runs pass every check}. Both need Python 3.11 or later and nothing beyond the standard library. The figures also need \code{matplotlib} (\code{python harness/figures.py}). On the author's laptop the two commands take about five seconds in total.

\subsection{What the archive cannot show}

The chains were read from PostgreSQL during each run and are not part of the archive. The archive therefore allows re-checking counts and verdicts, but not re-hashing the events.

\subsection{Source code}

The source code of the system, \sys{}, will be released under the Apache-2.0 licence with the product's first public release. Until then, the archive contains everything needed to reproduce the tables and figures of this paper from the stored results. Repeating the fault experiments themselves also needs the system's source, Docker and the images named in \appref{app:commands}, which also lists the commands.

%% file: sections/conclusion.tex
\section{Conclusion}
\label{sec:conclusion}

A tamper-evident record is evidence only if it is also complete. We stated completeness for an agent policy decision point as five checkable properties, and tested them with a client-side oracle under \NFaultTypes{} kinds of fault of the PDP's dependencies and of the PDP itself. No returned decision was missing from the chain, none was recorded twice, no record lacked a request, and every chain verified offline. A control with asynchronous commit showed that the oracle catches a loss that chain verification accepts.

What the faults did produce were records of decisions that the client never saw in time. These are in the safe direction, but they mean that a chain lists decisions, not executions. The price of the guarantee is a per-tenant ceiling, set by what one row lock holds: several database round trips, client-side work and a durable commit.

The next steps are shortening what the lock holds, propagating client deadlines and completing the idempotency record inside the append transaction. The oracle and the fault matrix released with this paper are the means of checking them. Independent replication on a Linux host with the database on a separate machine is the measurement that this study most needs.

%% file: sections/ack.tex
The author designed the study, ran and checked the experiments, and is responsible for all content. Generative AI assistants were used for drafting and coding support.

%% file: sections/glossary.tex
\section{Glossary of terms and abbreviations}
\label{app:glossary}

\begingroup
\setstretch{1.05}
\small
\begin{longtable}{@{}>{\raggedright\arraybackslash\bfseries}p{0.25\linewidth}>{\raggedright\arraybackslash}p{0.69\linewidth}@{}}
\caption{Terms and abbreviations used in this paper, in alphabetical order.}\label{tab:glossary}\\
\toprule
Term & Meaning\\
\midrule
\endfirsthead
\toprule
Term & Meaning\\
\midrule
\endhead
\bottomrule
\endfoot
\bottomrule
\endlastfoot
503 & HTTP status code ``Service Unavailable''. Here it is the PDP's answer when it cannot complete a decision and record it.\\
\addlinespace
Agent (AI agent) & A program, usually driven by a language model, that takes actions by calling tools.\\
\addlinespace
Append & Writing one event to a tenant's evidence chain: lock the chain head, check the event id, seal, insert, advance the head, commit.\\
\addlinespace
Asynchronous commit & The PostgreSQL setting \code{synchronous\_commit=off}. A commit returns before its WAL record is flushed, so a crash can lose transactions already reported as committed.\\
\addlinespace
Audit event & A record of one governed call: its input, the result of every rule, the policy version, the decision and the operation id.\\
\addlinespace
Chain head & The newest event of a tenant's chain. It is kept in one database row (\code{chain\_head}) that is locked during an append.\\
\addlinespace
Closed (fail mode) & See \emph{fail-closed}.\\
\addlinespace
Container & An isolated process environment (here Docker) in which a service runs.\\
\addlinespace
Coordinated omission & A measurement error in which a load generator slows down when the server stalls, so the stall is hidden from the latency figures. An open-loop generator avoids it.\\
\addlinespace
Coverage & The share of an agent's actions that actually pass through the governance layer. Assumed to be complete in this paper.\\
\addlinespace
Decision & The PDP's verdict on a tool call: \code{ALLOW}, \code{DENY} or \code{REQUIRE\_APPROVAL}.\\
\addlinespace
Durable commit & A database commit that returns only after its WAL record is flushed to disk (\code{synchronous\_commit=on}).\\
\addlinespace
Effective decision & The decision that takes effect after the mode is applied. In observe mode a policy \code{DENY} has effective decision \code{ALLOW}.\\
\addlinespace
Egress proxy & A proxy through which an agent's outbound network requests pass. It can ask the PDP before it forwards a request.\\
\addlinespace
Enforce (mode) & A mode in which the policy's decision takes effect.\\
\addlinespace
Enforcement point & The component that asks the PDP and acts on its answer. Here, the SDK.\\
\addlinespace
Evidence chain & A tenant's hash chain of audit events, stored in PostgreSQL.\\
\addlinespace
Fail mode & What the client does when it gets no answer from the PDP: \code{closed} (refuse the call) or \code{open\_with\_alert} (let it run and write a gap marker).\\
\addlinespace
Fail-closed, fail-open & A system that refuses to act when its control cannot decide is fail-closed. One that lets the action proceed is fail-open.\\
\addlinespace
Fault injection & Deliberately causing a failure (a crash, a freeze, a delay) to observe how a system behaves.\\
\addlinespace
fsync & A system call that forces data written to a file onto the disk before it returns.\\
\addlinespace
Gap marker & An \code{UNGOVERNED\_ACTION} record that the SDK writes to a local journal when a call runs without a decision from the PDP.\\
\addlinespace
Genesis value & The fixed starting hash $h_0$ of a tenant's chain.\\
\addlinespace
Governance layer & A control layer between an agent and its tools. It decides whether each tool call may run and records the decision.\\
\addlinespace
Hash chain & A sequence of records in which each stores a hash that covers itself and the previous record, so changing or removing an earlier record breaks all later hashes.\\
\addlinespace
HTTP & Hypertext Transfer Protocol, used here for the requests between the SDK and the PDP.\\
\addlinespace
Idempotency key & An identifier sent with a request so that a retry of the same request is recognised and answered from the stored first result instead of being processed again.\\
\addlinespace
IETF & Internet Engineering Task Force, the body that publishes internet standards (RFCs) and drafts.\\
\addlinespace
JCS & JSON Canonicalization Scheme (RFC 8785): a fixed way to write JSON so that equal data gives identical bytes.\\
\addlinespace
JSONL & A text file with one JSON document per line.\\
\addlinespace
Kill switch & A flag with which an operator can stop an agent or a tool at once.\\
\addlinespace
Late answer & An answer that reaches the client more than 2\,s after the request was sent, the SDK's time-out.\\
\addlinespace
LLM & Large language model.\\
\addlinespace
MCP & Model Context Protocol, a standard way for agents to call tools. An MCP gateway sits between an agent and its tool servers.\\
\addlinespace
Merkle tree, checkpoint & A binary tree of hashes that lets one prove with a short proof that an event belongs to a range. A Merkle checkpoint is a signed root over a range of events.\\
\addlinespace
Mode & A role's setting that says whether the policy's decision takes effect (\code{enforce}) or is only recorded (\code{observe}).\\
\addlinespace
NTFS & The Windows file system, used by the host's disk.\\
\addlinespace
Oathpost & The runtime governance layer for AI agents whose PDP is studied here.\\
\addlinespace
Observe (mode) & A mode in which a call that the policy would deny is recorded as \code{DENY} but runs (effective decision \code{ALLOW}).\\
\addlinespace
OPA & Open Policy Agent, a policy engine that evaluates rules written in the Rego language.\\
\addlinespace
Open-loop load & A load generator that releases requests on a fixed timetable regardless of how fast the server answers.\\
\addlinespace
\code{open\_with\_alert} & A fail mode: with no answer from the PDP, the client lets the call run and writes a gap marker. See \emph{fail-open}.\\
\addlinespace
Operation id & A unique identifier given to each request. It is stored in the audit event and used to match a request with its record.\\
\addlinespace
Oracle & Here, the checker that compares every request with the chain after a run and tests the five properties.\\
\addlinespace
p50, p99 & Percentiles of latency. p50 is the median; p99 is the value below which 99 of every 100 requests fall.\\
\addlinespace
PDP & Policy decision point: the service that evaluates a tool call against the policy and returns a decision.\\
\addlinespace
Phantom record & A decision event that belongs to no request that was sent.\\
\addlinespace
Policy, policy bundle & A policy is a set of rules written by an organisation. A bundle is a versioned package of the policy that a PDP serves.\\
\addlinespace
Positive control & A deliberately weakened system on which a checker must fire, used to show that the checker is not blind.\\
\addlinespace
PostgreSQL & A relational database. Here it stores the tool registry, grants, kill switches, approvals and the evidence chain. TimescaleDB is a PostgreSQL extension for time-series data; the evidence table is one of its tables.\\
\addlinespace
Prometheus & An open-source monitoring system. The PDP exposes counters and duration histograms in its format.\\
\addlinespace
Redis & An in-memory data store. Here it holds call counters and budgets.\\
\addlinespace
RFC & Request for Comments, a numbered IETF document.\\
\addlinespace
Role & The identity under which an agent acts, with its own mode and fail mode.\\
\addlinespace
SDK & Software development kit. Here, the client library that wraps a tool function so that each call first asks the PDP.\\
\addlinespace
SIGKILL & An operating-system signal that ends a process at once and cannot be handled by it, simulating a crash.\\
\addlinespace
SIGTERM & An operating-system signal that asks a process to shut down in an orderly way.\\
\addlinespace
TCP, RST & Transmission Control Protocol, the reliable byte stream beneath HTTP and the database connection. RST is the TCP segment that aborts a connection.\\
\addlinespace
Tenant & One customer organisation of the service. Each tenant has its own evidence chain.\\
\addlinespace
TLS & Transport Layer Security, the encryption of network connections.\\
\addlinespace
tmpfs & A memory-backed file system, used here to give the evidence table a small, fillable storage area.\\
\addlinespace
Tool call & A request by an agent to use a tool, for example to run a query or send an e-mail.\\
\addlinespace
uvicorn, FastAPI, psycopg & Software used by the PDP: a Python web server (uvicorn), a Python web framework (FastAPI) and a Python driver for PostgreSQL (psycopg).\\
\addlinespace
WAL & Write-ahead log: the log from which PostgreSQL recovers after a crash. A commit is durable once its WAL record is flushed to disk.\\
\addlinespace
WSL2 & Windows Subsystem for Linux 2: a lightweight virtual machine in which Linux containers run on Windows.\\
\end{longtable}
\endgroup

%% file: sections/appendix-commands.tex
\section{Commands to repeat the experiments}
\label{app:commands}

This appendix is for readers who want to repeat the fault experiments, not only to recompute the tables (\cref{sec:data}). It needs the system's source (location in the environment variable \code{OATHPOST\_REPO}), Python 3.13, Docker, and the images \code{timescale/timescaledb:latest-pg16}, \code{redis:7-alpine} and \code{openpolicyagent/opa:latest-static}. The commands run from the top directory of the archive.

\begin{small}
\begin{verbatim}
python harness/stack_p2.py up              # PostgreSQL and Redis containers
python harness/run_matrix.py --reps 5      # control and 12 fault types, x5
python harness/run_matrix.py --reps 5 --faults pg-kill --extra "--sync off"
python harness/diskfull_setup.py up        # 64 MB tmpfs tablespace
for r in 1 2 3 4 5; do
  python harness/run_fault.py --fault disk-full --rep $r \
      --state .p2-diskstate.json
done
python harness/diskfull_setup.py down
for f in none pg-kill pdp-kill redis-kill; do
  for r in 1 2 3 4 5; do
    python harness/sdk_failmode.py --fault $f --rep $r
  done
done
python harness/run_sweeps.py --reps 5
python harness/run_append.py --reps 5
python harness/jsonl_baseline.py
python harness/analyze.py                  # tables and paper/generated
python harness/independent_check.py        # re-check of the oracle outputs
python harness/stack_p2.py down
\end{verbatim}
\end{small}

\noindent The thirteenth fault type, the full evidence storage, is run by the separate loop above because it needs its own database with the tablespace.